\documentclass[aps,prd,amsmath,floats,floatfix, twocolumn,
superscriptaddress,nofootinbib,showpacs,longbibliography]{revtex4-1}

\usepackage[colorlinks, pdfborder={0 0 0}, plainpages=false]{hyperref}
\usepackage{graphicx}
\usepackage{xspace}
\usepackage[usenames,dvipsnames]{color}
\usepackage{amssymb}
\usepackage[normalem]{ulem} 
\usepackage[all]{hypcap}

\DeclareMathAlphabet{\mathpzc}{OT1}{pzc}{m}{it}

\newcommand{\roughly}{\mathchar"5218\relax} 

\newcommand{\h}{\mathpzc{h}}
\newcommand{\hlm}{\mathpzc{h}_{\ell m}}
\newcommand{\Alm}{A_{\ell m}}
\newcommand{\omegalm}{\omega_{\ell m}}
\newcommand{\philm}{\phi_{\ell m}}
\newcommand{\chieff}{\chi_{\mathrm{eff}}}

\newcommand{\ins}{\mathrm{ins}}

\newcommand{\AEI}{\affiliation{Max Planck Institute for Gravitational Physics
    (Albert Einstein Institute), Am M\"uhlenberg 1, Potsdam 14476, Germany}} %
\newcommand{\Caltech}{\affiliation{Theoretical Astrophysics,
    California Institute of Technology, Pasadena, CA 91125, USA}}
\newcommand{\Cornell}{\affiliation{Center for Radiophysics and Space
    Research, Cornell University, Ithaca, New York 14853, USA}}
\newcommand{\UMassD}{\affiliation{Department of Mathematics,
    Center for Scientific Computing and Visualization Research,
    University of Massachusetts, Dartmouth, MA 02747, USA}}

\begin{document}

\title{
Surrogate model of hybridized numerical relativity binary black hole waveforms
}

\author{Vijay Varma} \Caltech

\author{Scott E. Field} \UMassD

\author{Mark A. Scheel} \Caltech

\author{Jonathan Blackman} \Caltech

\author{Lawrence E. Kidder} \Cornell

\author{Harald P. Pfeiffer} \AEI

\date{\today}

\begin{abstract}
Numerical relativity (NR) simulations provide the most accurate binary black
hole gravitational waveforms, but are prohibitively expensive for applications
such as parameter estimation. Surrogate models of NR waveforms have been shown
to be both fast and accurate.  However, NR-based surrogate models are limited
by the training waveforms' length, which is typically about 20 orbits before
merger.  We remedy this by hybridizing the NR waveforms using both
post-Newtonian and effective one body waveforms for the early inspiral.
We present NRHybSur3dq8, a surrogate model for hybridized nonprecessing
numerical relativity waveforms, that is valid for the entire LIGO band
(starting at $20~\text{Hz}$) for stellar mass binaries with total masses as low
as $2.25\,M_{\odot}$.  We include the $\ell \leq 4$ and $(5,5)$ spin-weighted
spherical harmonic modes but not the $(4,1)$ or $(4,0)$ modes.  This model has
been trained against hybridized waveforms based on 104 NR waveforms with mass
ratios $q\leq8$, and $|\chi_{1z}|,|\chi_{2z}| \leq 0.8$, where $\chi_{1z}$
($\chi_{2z}$) is the spin of the heavier (lighter) BH in the direction of
orbital angular momentum.  The surrogate reproduces the hybrid waveforms
accurately, with mismatches $\lesssim 3\times10^{-4}$ over the mass range
$2.25M_{\odot} \leq M \leq 300 M_{\odot}$.  At high masses
($M\gtrsim40M_{\odot}$), where the merger and ringdown are
more prominent, we show
roughly two orders of magnitude improvement over existing waveform models.  We
also show that the surrogate works well even when extrapolated outside its
training parameter space range, including at spins as large as 0.998.  Finally,
we show that this model accurately reproduces the spheroidal-spherical mode
mixing present in the NR ringdown signal.
\end{abstract}


\maketitle

\section{Introduction}
The era of gravitational wave (GW) astronomy has been emphatically unveiled
with the recent detections~\cite{LIGOVirgo2016a, TheLIGOScientific:2017qsa,
Abbott:2016nmj, Abbott:2017vtc, Abbott:2017gyy, Abbott:2017oio,
LIGOVirgo2018:GWTC_1} by LIGO~\cite{aLIGO2} and Virgo~\cite{aVirgo2}.
The detection of gravitational wave signals from compact binary sources is expected to become a
routine occurrence as the advanced detectors reach their design
sensitivity~\cite{Abbott:2016wya, LIGOScientific:2018jsj}. The possible science
output from these events crucially depends on the availability of an accurate
waveform model to compare against observed signals.

Numerical relativity (NR) is the only {\it ab initio} approach that accurately
produces waveforms from the merger of a binary black hole (BBH) system.
However, because NR simulations are computationally expensive, it is
impractical to use them directly for applications such as parameter estimation,
which can require upwards of $10^7$ waveform evaluations.  Therefore, the GW
community has developed several approximate waveform
models~\cite{Khan:2018fmp, Pan:2013rra, London:2017bcn, Cotesta:2018fcv,
Khan:2015jqa, Bohe:2016gbl, Hannam:2013oca, Taracchini:2013rva, PanEtAl:2011,
Mehta:2017jpq}, some of which are fast to evaluate.
These models make certain physically-motivated assumptions
about the underlying phenomenology of the waveforms, and they fit for any
remaining free parameters using NR simulations.

Surrogate modeling~\cite{Field:2013cfa,Purrer:2014fza} is an alternative
approach that doesn't assume an underlying phenomenology and has been applied
to a diverse range of problems~\cite{Field:2013cfa, Purrer:2014fza,
    Lackey:2016krb, Doctor:2017csx, Blackman:2015pia, Blackman:2017dfb,
Blackman:2017pcm, Huerta:2017kez, Chua:2018woh, Canizares:2014fya,
Galley:2016mvy}. NR Surrogate models follow a data-driven approach, directly
using the NR waveforms to implicitly reconstruct the underlying phenomenology.
Three NR surrogate models have been built so far~\cite{Blackman:2015pia,
Blackman:2017dfb, Blackman:2017pcm}, including a 7-dimensional (mass ratio $q$
and two spin vectors) model for generically precessing systems in
quasi-circular orbit~\cite{Blackman:2017pcm}.  Through cross-validation
studies, these models were shown to be nearly as accurate as the NR waveforms
they were trained against.

Despite the success of the surrogate modeling approach, existing surrogate
models have two important limitations: (1) Because they are based solely on NR
simulations, which typically are only able to cover the last $\roughly 20$
orbits of a BBH inspiral, they are not long enough to span the full LIGO band
for stellar mass binaries. (2) Apart from the first non-spinning
model~\cite{Blackman:2015pia}, these models have been restricted to mass ratios
$q\leq2$~\footnote{We use the convention $q=m_1/m_2$, where $m_1$ and
$m_2$ are the masses of the component black holes, with $m_1\ge m_2$.}. There
are two reasons for this: (i) The 7d parameter space is vast, requiring at
least a few thousand simulations to sufficiently cover it.  (ii) Because of the
smaller length scale introduced by the lighter black hole, NR simulations
become increasingly more expensive with mass ratio.

In this work we address these limitations in the context of nonprecessing BBH
systems. First, to include the early inspiral we ``hybridize'' the NR waveforms
: each full waveform consists of a post-Newtonian (PN) and effective one body
(EOB) waveform at early times that is smoothly attached to an NR waveform at
late times. Second, since we restrict ourselves to the 3-dimensional space of
nonprecessing BBHs, fewer simulations are necessary compared to the
7-dimensional case, and therefore we can direct computational resources to
simulations with higher mass ratios. The resulting model, NRHybSur3dq8, is the
first NR-based surrogate model to span the entire LIGO frequency band for
stellar mass binaries; assuming a detector low-frequency cut-off of $20
~\text{Hz}$, this model is valid for total masses as low as $2.25 M_{\odot}$.
This model is based on 104 NR waveforms in the parameter range $q\leq8$, and
$|\chi_{1z}|,|\chi_{2z}|\leq0.8$, where $\chi_{1z}$ ($\chi_{2z}$) is the dimensionless spin
of the heavier (lighter) black hole (BH).

The plus ($h_{+}$) and cross ($h_{\times}$) polarizations of GWs can be
conveniently represented by a single complex time-series, $\h = h_{+} -i
h_{\times}$. The complex waveform on a sphere can be decomposed into a sum of
spin-weighted spherical harmonic modes $\hlm$~\cite{NewmanPenrose1966,
  Goldberg1967}, so that the
waveform along any direction ($\iota$,$\varphi_0$) in the binary's
source frame is given by
\begin{gather}
    \h(t, \iota, \varphi_0) = \sum^{\infty}_{\ell=2} \sum_{m=-l}^{l}
        \hlm(t) ~^{-2}Y_{\ell m}(\iota, \varphi_0),
\label{Eq:spherical_harm}
\end{gather}
where $^{-2}Y_{\ell m}$ are the spin$\,=\!\!-2$ weighted spherical harmonics,
$\iota$ is the inclination angle between the orbital angular momentum of the
binary and line-of-sight to the detector, and $\varphi_0$ is the initial binary
phase.  $\varphi_0$ can also be thought of as the azimuthal angle
between the $x-$axis of the source frame and the line-of-sight to the detector.
We define the source frame as follows: The $z-$axis is along the orbital
angular momentum direction, which is constant for nonprecessing BBH. The
$x-$axis is along the line of separation from the lighter BH to the heavier BH
at some reference time/frequency. The $y-$axis completes the triad.

The $\ell \!\!=\!\! |m| \!\!=\!\! 2$ terms typically dominate the sum in
Eq.~(\ref{Eq:spherical_harm}), and are referred to as the \emph{quadrupole}
modes. Studies~\cite{Varma:2016dnf, Capano:2013raa, Littenberg:2012uj,
Bustillo:2016gid, Brown:2012nn, Varma:2014, Graff:2015bba, Harry:2017weg} have
shown that the nonquadrupole modes, while being subdominant, can play a
nonnegligible role in detection and parameter estimation of GW sources,
particularly for large signal to noise ratio (SNR), large total mass, large mass
ratio, or large inclination angle $\iota$.  For the first event,
GW150914~\cite{LIGOVirgo2016a}, the systematic errors due to the
quadrupole-mode-only approximation are generally smaller than the statistical
errors~\cite{Abbott:2016wiq, Abbott:2016apu}, although higher modes may lead to
modest changes in some of the extrinsic parameter values~\cite{Kumar:2018hml}.
However, as the detectors approach their design
sensitivity~\cite{Abbott:2016wya}, one should prepare for high-SNR sources
(particularly at larger mass ratios than those seen so far), where the
quadrupole-mode-only approximation breaks down. In addition, nonquadrupole modes
can help break the degeneracy between the binary inclination and distance,
which is present for quadrupole-mode-only models (see
e.g.~\cite{London:2017bcn, OShaughnessy:2014shr, Usman:2018imj}).

In this work, we model the following spin-weighted spherical harmonic modes:
$\ell \leq 4$ and (5,5), but not the (4,1) or (4,0) modes~\footnote{Because of
the symmetries of nonprecessing BBHs (see Eq.~(\ref{Eq:negm_symmetry})), the
$m<0$ modes contain the same information as the $m>0$ modes, and do not need to
be modeled separately.}.  Several inspiral-merger-ringdown waveform
models~\cite{Cotesta:2018fcv, London:2017bcn, PanEtAl:2011, Mehta:2017jpq} that
include nonquadrupole modes have been developed in recent years; however,
compared to those models we show an improved accuracy and we include more
modes.

The rest of the paper is organized as follows. In Sec.~\ref{Sec:TrainingSet} we
choose the parameters at which to perform NR simulations, which will be used
for training the surrogate model. Sec.~\ref{Sec:NR_runs} describes the NR
simulations. Sec.~\ref{Sec:inspiral_waveforms} describes our procedure to
compute the waveform for the early inspiral using PN and EOB waveforms.
Sec.~\ref{Sec:Hybridization} describes our hybridization procedure to attach
the early inspiral waveform to the NR waveforms. Sec.~\ref{Sec:buildSurrogate}
describes the construction of the surrogate model. In Sec.~\ref{Sec:Results},
we test the surrogate model by comparing against NR and hybrid waveforms. We
end with some concluding remarks in Sec.~\ref{Sec:Conclusion}.  We make our
model available publicly through the easy-to-use Python package
\emph{gwsurrogate}~\cite{gwsurrogate}. In addition, our model is implemented in
C with Python wrapping in the LIGO Algorithm Library~\cite{lalsuite}. We
provide an example Python evaluation code at \cite{SpECSurrogates}.

\section{Training set generation}
\label{Sec:TrainingSet}

\subsection{Greedy parameters from PN surrogate model}
We do not know a priori the distribution or number of NR simulations required
to build an accurate surrogate model. Furthermore, we hope to select a
representative distribution that will allow for an accurate surrogate to be
built with as few NR simulations as possible. Therefore, we estimate this
distribution by first building a surrogate model for PN waveforms; we find that
parameters suitable for building an accurate PN surrogate are also suitable for
building an NR or a hybrid NR-PN surrogate.

We use the same methods to build the PN surrogate as we use for the hybrid
surrogate (cf. Sec.~\ref{Sec:buildSurrogate}). We use the PN waveforms
described in Sec.~\ref{Subsec:pn_waveforms}; however, for simplicity we only
model the (2,2) mode. In addition, we restrict the length of the PN waveforms
to be $5000M$, terminating at the innermost-stable-circular-orbit's orbital
frequency, $\omega_{\mathrm{orb}}\!=\!{6^{-3/2}}~\text{rad}/M$, where $M$ is
the total mass of the binary.

We determine the desired training data set of parameters as follows. We begin
with just the corner cases of the parameter space; for the 3d case considered
here, that consists of 8 points at $(q,\chi_{1z},\chi_{2z}) = (1 ~\text{or}~ 8,
\pm0.8, \pm0.8)$.  We build up the desired set of parameters iteratively, in a
greedy manner: At each iteration we build a PN surrogate using the current
training data set and test the model against a much larger ($\sim 10$ times)
validation data set. The validation data set is generated by randomly
resampling the parameter space at each iteration. Since the boundary cases are
expected to be more important, for 30\% of the points in the validation set we
sample only from the boundary of the parameter space, which corresponds to the
faces of a cube in the 3d case.  We select the parameter in the validation set
that has the largest error (cf.  Eq.~\eqref{Eq:time_dom_mismatch}), and add
this to our training set (hence the name greedy parameters).  We repeat until
the validation error reaches a certain threshold.

In order to estimate the difference between two complexified waveforms, $\h_1$
and $\h_2$, we use the time-domain mismatch,
\begin{gather}
\label{Eq:time_dom_mismatch}
\mathcal{MM} = 1 - \frac{\left<\h_1, \h_2\right>}{\sqrt{\left<\h_1, \h_1\right>
    \left<\h_2, \h_2\right>}}, \\
\label{Eq:time_dom_mismatch_2}
\left<\h_1, \h_2\right> = \left| \int_{t_{\mathrm{min}}}^{t_{\mathrm{max}}}
    \h_1 (t) {\h}_2^* (t) dt \right|,
\end{gather}
where $*$ indicates a complex conjugation, and $|.|$ indicates the absolute
value.  Note that in this section, we do not perform an optimization over time
and phase shifts. In addition, we assume a flat noise curve.

\begin{figure}[thb]
\includegraphics[width=0.475\textwidth]{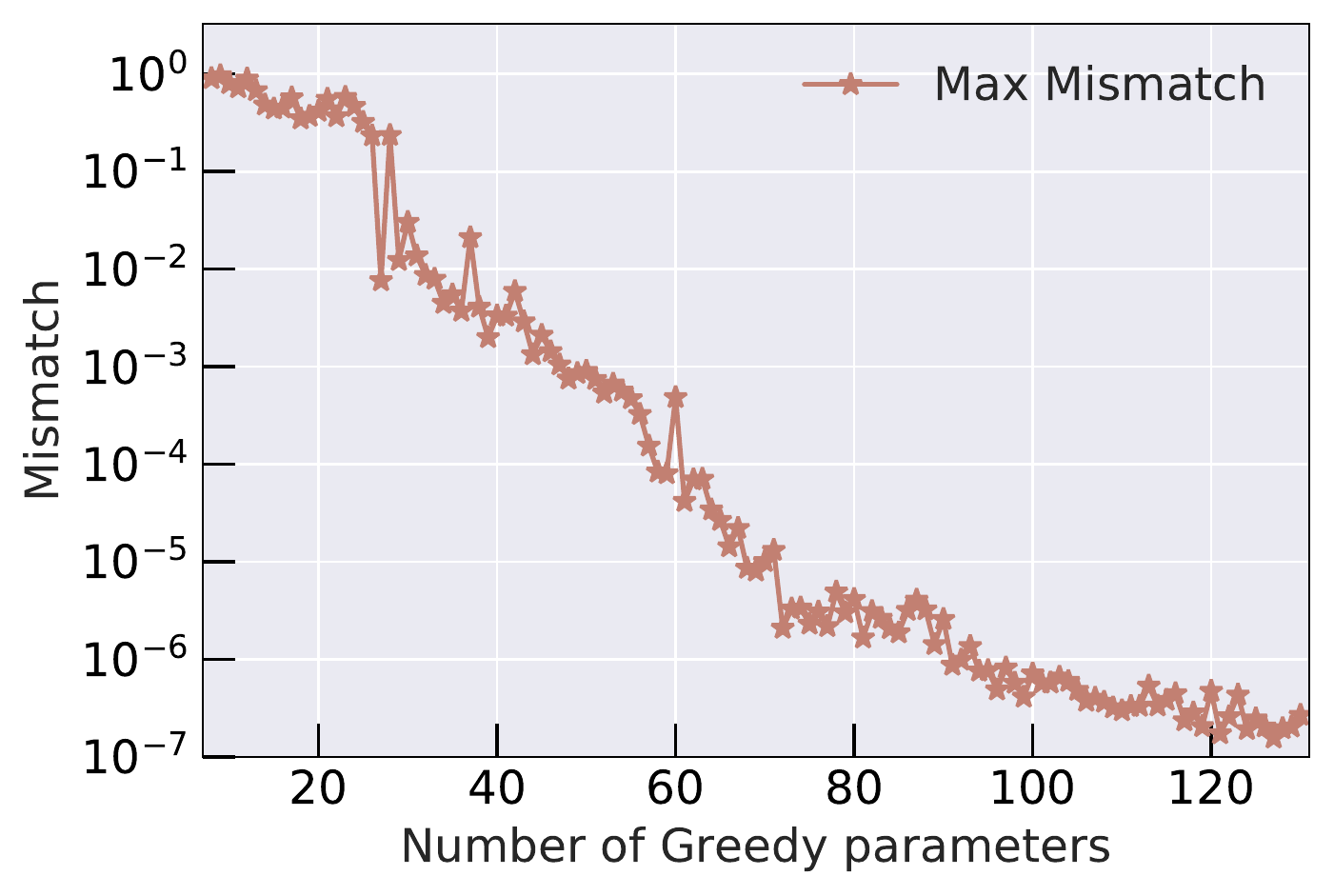}
\caption{Largest mismatch of the surrogate (over the entire validation set) as
a function of number of greedy parameters used to train the PN surrogate.
The PN surrogate is seen to converge to the validation waveforms as the size of
the training data set increases.
}
\label{Fig:PNGreedyParams}
\end{figure}

Figure~\ref{Fig:PNGreedyParams} shows how the maximum validation error
decreases as we add greedy parameters to our training data set. For our case,
we stop at 100 greedy parameters (at which point the mismatch is $<10^{-6}$)
and use those parameters to perform the NR simulations. Note that we don't
expect 100 NR simulations to produce an NR surrogate with comparable accuracy,
$\mathcal{MM}<10^{-6}$, for two reasons.  First, unlike the PN waveforms used
here, the NR simulations also include the merger-ringdown part, which we expect
to be more difficult to model. Second, the NR numerical truncation error is typically
higher than $10^{-6}$ in mismatch, therefore the numerical noise will limit the
accuracy.

\section{NR simulations}
\label{Sec:NR_runs}

\begin{figure}[thb]
\includegraphics[width=0.5\textwidth]{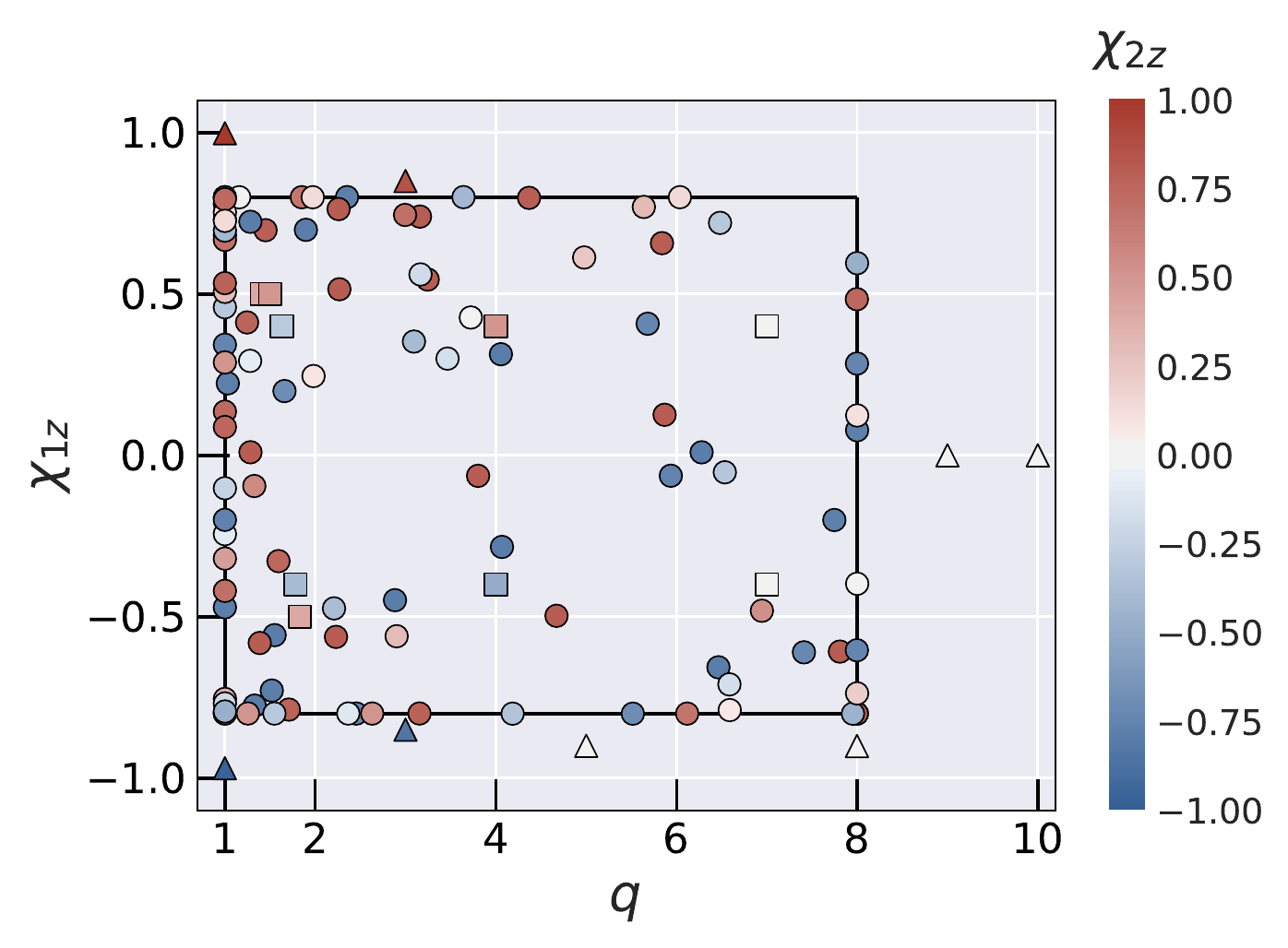}
\caption{The parameter space covered by the 104 NR waveforms (circle markers)
used in the construction of the surrogate model in
Sec.~\ref{Sec:buildSurrogate}. We also show the 9 long NR waveforms (square
markers) used to test hybridization in Sec.~\ref{Subsec:longNRTests}, and the 8
NR waveforms (triangle markers) used to test extrapolation in
Sec.~\ref{Subsec:extrapTests}. The axes show the mass ratio and the spin on the
heavier BH, while the colors indicate the spin on the lighter BH. The black
rectangle indicates the bounds of the training region: $1\leq q \leq 8$, $-0.8
\leq \chi_{1z},\chi_{2z} \leq 0.8$.
}
\label{Fig:NRParams}
\end{figure}

The NR simulations for this model are performed using the Spectral Einstein
Code (SpEC)~\cite{SpECwebsite, Pfeiffer2003, Lovelace2008, Lindblom2006,
Szilagyi:2009qz, Scheel2009} developed by the SXS~\cite{SXSWebsite}
collaboration. Of the 100 cases determined in Sec.~\ref{Sec:TrainingSet}, only
91 simulations were successfully completed\footnote{The main reason for failure
is large constraint violation as the binary approaches merger. We believe a
better gauge condition may be needed for some of these simulations.}. These
simulations have been assigned the identifiers SXS:BBH:1419 - SXS:BBH:1509, and
are made publicly available through the SXS public catalog~\cite{SXSCatalog}.
For cases with equal mass, but unequal spins, we can exchange the two BHs to
get an extra data point.  There are 13 such cases, leading to a total of 104 NR
waveforms.  These are shown as circular markers in Fig.~\ref{Fig:NRParams}.

The start time of these simulations varies between $4270 M$ and $5227 M$ before
the peak of the waveform amplitude (defined in Eq.~(\ref{Eq:total_amplitude})),
where $M = m_1 + m_2$ is the total Christodoulou mass measured after the
initial burst of junk radiation. The algorithm for choosing a fiducial time at
which junk radiation ends is discussed in Ref.~\cite{SXSCatalog2018}.  The
initial orbital parameters are chosen through an iterative
procedure~\cite{Buonanno:2010yk} such that the orbits are quasicircular; the
largest eccentricity for these simulations is $7.5 \times 10^{-4}$, while the
median value is $4.2 \times 10^{-4}$.  The waveforms are extracted at several
extraction surfaces at varying finite radii form the origin and then
extrapolated to future null infinity~\cite{Boyle-Mroue:2008}.  Finally, the
extrapolated waveforms are corrected to account for the initial drift of the
center of mass~\cite{Boyle2015a, scri}. The time steps during the simulations
are chosen nonuniformly using an adaptive time-stepper~\cite{SXSCatalog2018}.
We interpolate these data to a uniform time step of $0.1 M$; this is dense
enough to capture all frequencies of interest, including near merger.

\section{Early inspiral waveforms}
\label{Sec:inspiral_waveforms}

While NR provides accurate waveforms, computational constraints limit NR to
only the late inspiral, merger, and ringdown phases.  Fortunately, PN/EOB
waveforms are expected to be accurate in the early inspiral. Hence we can
``stitch'' together an early inspiral waveform and an NR waveform, to get a
\emph{hybrid} waveform~\cite{Santamaria:2010yb, Ohme:2011rm, Ohme:2011zm,
MacDonald:2011ne, MacDonald:2012mp, Ajith:2012az, Varma:2014, Boyle:2011dy,
Bustillo:2015ova} that spans the entire frequency range relevant for
ground-based detectors. In this section, we describe the waveforms we use for
the early inspiral, leaving the hybridization procedure for the next section.

\subsection{PN waveforms}
\label{Subsec:pn_waveforms}

We first generate PN waveforms as implemented in the GWFrames
package~\cite{gwframes}. For the orbital phase we include nonspinning terms up
to 4 PN order~\cite{Blanchet04a, Blanchet2014, Jaranowski:2013lca,
BiniDamour:2013, Bini:2013rfa} and spin terms up to 2.5 PN
order~\cite{Kidder:1995zr, Will96, Bohe:2012mr}. We use the
TaylorT4~\cite{Boyle2007} approximant to generate the PN phase; however, as
described below, we replace this phase with an EOB-derived phase. For the
amplitudes, we include terms up to 3.5 PN order~\cite{BFIS, FayeEtAl:2012,
FayeEtAl:2014}.

The spherical harmonic modes of the PN waveform can be written (after rescaling
to unit total mass and unit distance) as~\cite{BFIS, Blanchet2014},
\begin{gather}
    \hlm^\mathrm{PN} = 2 ~\eta ~(v^\mathrm{PN})^2 ~\sqrt{\frac{16 \pi}{5}}
    H_{\ell m}^\mathrm{PN} e^{-\mathrm{i} m \phi_{\mathrm{orb}}^\mathrm{PN}},
\end{gather}
where $\eta=q/(1+q)^2$ is the symmetric mass ratio, $v^\mathrm{PN}$ is the
characteristic speed that sets the perturbation scale in PN,
$\phi^\mathrm{PN}_\mathrm{orb}$ is the (real) orbital phase, and $H_{\ell
m}^\mathrm{PN}$ are the complex amplitudes of different modes. Note that we
ignore the tail distortions~\cite{Blanchet93, Arun:2004} to the orbital phase
as these are 4 PN corrections (see e.g.~\cite{Inprep-Barkett:2018}).

The complex strain $\hlm^\mathrm{PN}$ is obtained as a time series from
GWFrames. We can absorb the complex part of the amplitudes into the phases and
rewrite the strain as
\begin{gather}
    \label{Eq:pn_modes}
    \hlm^\mathrm{PN} =  A_{\ell m}^\mathrm{PN} e^{-\mathrm{i} \phi_{\ell m}^\mathrm{PN}}, \\
    \label{Eq:pn_phase}
    \phi_{\ell m}^\mathrm{PN} = m ~\phi_\mathrm{orb}^\mathrm{PN} + \xi^\mathrm{PN}_{\ell m} , \\
    \label{Eq:pn_orbphase}
    \phi_\mathrm{orb}^\mathrm{PN} = \frac{\phi_{22}^\mathrm{PN}}{2},
\end{gather}
where $A_{\ell m}^\mathrm{PN}$ and $\phi_{\ell m}^\mathrm{PN}$ are the real
amplitude and phase of a given mode, and $\xi^\mathrm{PN}_{\ell m}$ is an
offset that captures the complex part of $H_{\ell m}^\mathrm{PN}$. Note that
Eqs.~(\ref{Eq:pn_phase}) and (\ref{Eq:pn_orbphase}) together imply
$\xi^\mathrm{PN}_{22}=0$; $H_{22}^\mathrm{PN}$ contains complex terms starting
at 2.5PN, but these appear as 5PN corrections in the phase (see
e.g.~\cite{Inprep-Barkett:2018}), which we can safely ignore.

At this stage, $A_{\ell m}^\mathrm{PN}$, $\phi_{\ell m}^\mathrm{PN}$, and
$\xi^\mathrm{PN}_{\ell m}$
are functions of time. But they can be recast as
functions of the characteristic speed by
first computing
\begin{gather}
    \label{Eq:pn_char_speed}
    v^\mathrm{PN}(t) =
    \left(\frac{d \phi^\mathrm{PN}_\mathrm{orb}}{dt} \right)^{1/3},
\end{gather}
where the derivative is performed numerically, and then inverting
Eq.~(\ref{Eq:pn_char_speed})
to obtain $t(v^\mathrm{PN})$.  Then we define
\begin{gather}
    \label{Eq:pn_amplitude}
    A_{\ell m}^\mathrm{PN}(v^\mathrm{PN}) = |\h_{\ell m}^\mathrm{PN}(t(v^\mathrm{PN}))|, \\
    \xi^\mathrm{PN}_{\ell m}(v^\mathrm{PN})
        = \phi_{\ell m}^\mathrm{PN}(t(v^\mathrm{PN}))- m ~\phi_\mathrm{orb}^\mathrm{PN}(t(v^\mathrm{PN})).
    \label{Eq:pn_phase_offset}
\end{gather}

Note that the PN waveform is generated in the source frame defined
such that
the reference time is the initial time. This also ensures that the
heavier BH is on the positive $x-$axis at the initial time, and the initial
orbital phase is zero.

To summarize: From the GWFrames package, we obtain the complex time series
$\h^\mathrm{PN}_{\ell m}$ (Eq.~(\ref{Eq:pn_modes})). We compute the orbital
phase (Eq.~(\ref{Eq:pn_orbphase})), the real amplitudes
(Eq.~(\ref{Eq:pn_amplitude})), and the phase offsets
(Eq.~(\ref{Eq:pn_phase_offset})).  These three quantities are obtained as a
time series but can be represented as functions of the characteristic speed
using Eq.~(\ref{Eq:pn_char_speed}).

\begin{figure*}[thb]
\includegraphics[width=\textwidth]{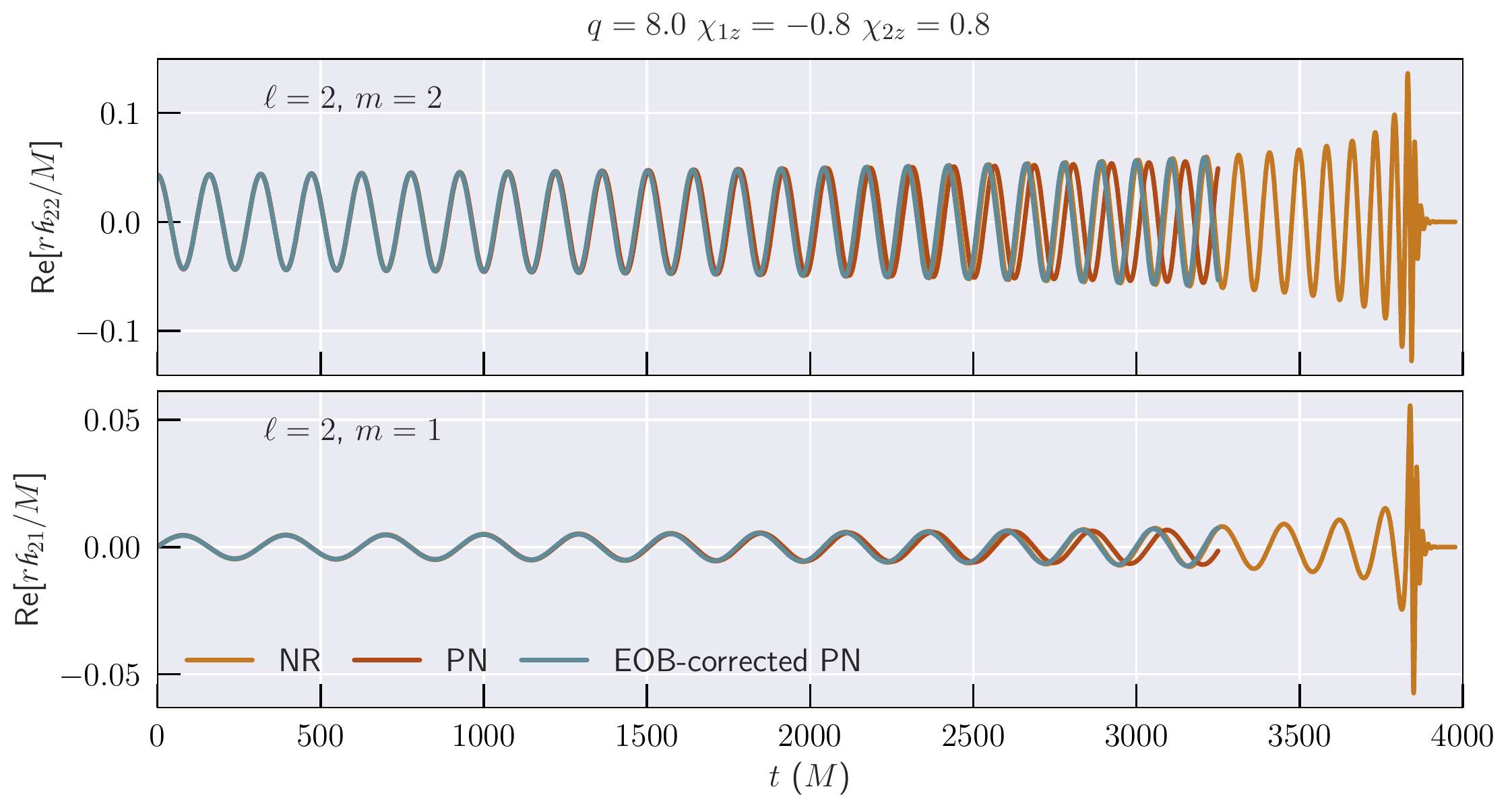}
\caption{
NR, PN (Sec.~\ref{Subsec:pn_waveforms}), and EOB-corrected PN
(Sec.~\ref{Subsec:EOB_correction}) waveforms for an example case. We show the
$(2,2)$ and $(2,1)$ modes. The binary parameters are shown at the top of the
plot. The EOB-corrected PN waveform~\cite{Varma:2014, Varma:2016dnf} stays
faithful to the NR waveform until much later times, compared to the pure PN
waveform.
}
\label{Fig:EOB_correction}
\end{figure*}

\subsection{EOB correction}
\label{Subsec:EOB_correction}

As was shown in previous works~\cite{Varma:2014, Varma:2016dnf}, we find that
the accuracy of the inspiral waveform can be improved by replacing the PN phase
with the phase derived from an NR-calibrated EOB model. For this work we use
SEOBNRv4~\cite{Bohe:2016gbl}.

SEOBNRv4 is a time domain model that includes only the
$(2,2)$ mode, which we can decompose as follows:
\begin{gather}
    \h^\mathrm{EOB}_{22} = A^\mathrm{EOB}_{22}~ e^{-\mathrm{i} \phi^\mathrm{EOB}_{22}},
\end{gather}
where $A^\mathrm{EOB}_{22}$ and $\phi^\mathrm{EOB}_{22}$ are the real amplitude
and phase of the $(2,2)$ mode. These are functions of time, but following the
same procedure as earlier, they can be recast in terms of the characteristic
speed:
\begin{gather}
    \label{Eq:eob_orbphase}
    \phi^\mathrm{EOB}_\mathrm{orb}(t) = \frac{\phi^\mathrm{EOB}_{22}(t)}{2}, \\
    v^\mathrm{EOB}(t) =  \left(\frac{d \phi^\mathrm{EOB}_\mathrm{orb}}{dt} \right)^{1/3},
    \label{Eq:eob_char_speed}
\end{gather}
where the derivative is performed numerically, and we invert
Eq.~(\ref{Eq:eob_char_speed}) to obtain $t(v^\mathrm{EOB})$.  We replace
$v^\mathrm{PN} \rightarrow v^\mathrm{EOB}$ in Eqs.~(\ref{Eq:pn_amplitude}) and
(\ref{Eq:pn_phase_offset}) to get, respectively, the EOB-corrected amplitudes
and phase offsets:
\begin{gather}
    A_{\ell m}^{\ins}(t) = A_{\ell m}^\mathrm{PN} (v^\mathrm{EOB}(t)), \\
    \xi^{\ins}_{\ell m}(t) = \xi^\mathrm{PN}_{\ell m} (v^\mathrm{EOB}(t)).
\end{gather}
Note that in practice, computing $A_{\ell m}^{\ins}(t)$ and $\xi^{\ins}_{\ell
m}(t)$ is accomplished via an interpolation in $v$: $A_{\ell m}^\mathrm{PN}(v)$
and $\xi^\mathrm{PN}_{\ell m}(v)$ as computed in Eqs.~(\ref{Eq:pn_amplitude})
and~(\ref{Eq:pn_phase_offset}) are known only at particular values of $v$,
which are $v^\mathrm{PN}(t_{i_\mathrm{PN}})$ where $t_{i_\mathrm{PN}}$ are the
times in the PN time series; we interpolate $A_{\ell m}^\mathrm{PN}(v)$ and
$\xi^\mathrm{PN}_{\ell m}(v)$ to the points
$v^\mathrm{EOB}(t_{i_\mathrm{EOB}})$ where $t_{i_\mathrm{EOB}}$ are the times
in the EOB time series.  We use a cubic-spline interpolation scheme as
implemented in Scipy~\cite{scipy}.

Following Eq.~(\ref{Eq:pn_phase}), the EOB-corrected phases are given by:
\begin{gather}
    \phi_{\ell m}^{\ins} =  m ~\phi_\mathrm{orb}^\mathrm{EOB} + \xi^{\ins}_{\ell m},
\end{gather}
where we use the EOB orbital phase from Eq.~(\ref{Eq:eob_orbphase}). Finally,
our EOB-corrected inspiral waveform modes are given by:
\begin{gather}
    \hlm^{\ins} =  A_{\ell m}^{\ins}~ e^{-\mathrm{i} \phi_{\ell m}^{\ins}}.
\end{gather}

Fig.~\ref{Fig:EOB_correction} shows an example of PN and EOB-corrected
waveforms along with the corresponding NR waveform. All three waveforms have
the same starting orbital frequency and their initial orbital phase is set to
zero. We see that the PN waveform becomes inaccurate at late times, as
expected. The EOB-corrected waveform, on the other hand, remains faithful to
the NR waveform until much later times.

\section{Hybridization}
\label{Sec:Hybridization}

In this section we describe our procedure to ``stitch'' together an inspiral
waveform (described in Sec.~\ref{Sec:inspiral_waveforms}) to an NR waveform
(described in Sec.~\ref{Sec:NR_runs}).

We start by generating inspiral and NR waveforms with the same component masses
and spins. We note that the spins measured in SpEC simulations agree well with
PN theory~\cite{Ossokine:2015vda}. However, the PN and NR waveforms are
typically represented in different coordinate systems that need to be aligned
with each other as follows.  The two coordinate systems are related to each
other by a
possible time translation and a possible rotation by three Euler angles:
inclination angle $\iota$, initial binary phase $\varphi_0$, and polarization
angle $\psi$.  For nonprecessing BBH the first angle $\iota$, is trivially
specified by requiring that the z-axis is along the direction of orbital
angular momentum.  This leaves us with the freedom to vary $\varphi_0$ and
$\psi$.  We choose the hybridization frame and time shifts by minimizing a cost
function in a suitable matching region; this is described in more detail below.

\subsection{Choice of cost function}

We use the following cost function when comparing two waveforms, $\mathpzc{h}$
and $\mathpzc{\tilde{h}}$, in the matching region:
\begin{gather}
\label{Eq:cost_function}
\mathcal{E} [\mathpzc{h} , \mathpzc{\tilde{h}}] =
    \frac{1}{2} \frac{\sum_{\ell,m}\int_{t_1}^{t_2}|
    \mathpzc{h}_{\ell m}(t) - \mathpzc{\tilde{h}}_{\ell m}(t)|^2 dt}
    {\sum_{\ell, m} \int_{t_1}^{t_2} |\mathpzc{h}_{lm}(t)|^2 dt},
\end{gather}
where $t_1$ and $t_2$ denote the start and end of the matching region, to be
defined in Sec.~\ref{Subsec:matching_region}, and the sum does not include
$m=0$ modes for reasons described in Sec.~\ref{Subsec:m0_modes}.  This cost
function was introduced in Ref.~\cite{Blackman:2017dfb} and is shown to be
related to the weighted average of the mismatch over the sky.

We minimize the cost function by varying the time and frame shifts between the
NR and inspiral waveforms
\begin{gather}
\label{Eq:frame_optimization}
\min_{t_0, \varphi_0, \psi} \mathcal{E}[
    \mathpzc{h}^\mathrm{NR}(t;\varphi_0, \psi), \mathpzc{h}^{\ins}(t; t_0)], \\
\label{Eq:shifted_nr}
\hlm^\mathrm{NR}(t; \varphi_0, \psi) =
    \hlm^\mathrm{NR}(t) ~e^{\mathrm{i} m \varphi_0} ~e^{2 \mathrm{i} \psi}, \\
\label{Eq:shifted_ins}
\hlm^{\ins}(t; t_0) = \hlm^{\ins}(t-t_0).
\end{gather}
We perform the time shifts on the inspiral waveform so that the matching region
always corresponds to the same segment of the NR waveform.  The frame shifts
are performed on the NR waveform so as to preserve the initial frame alignment
of the inspiral waveform (cf. Sec.~\ref{Subsec:pn_waveforms}). This alignment
gets inherited by the hybrid waveform, and is important in the surrogate
construction.

\subsection{$m=0$ modes}
\label{Subsec:m0_modes}

We find that the $m=0$ modes of the inspiral waveforms do not agree very well
with the NR waveforms. There are several possible reasons for
this~\cite{Favata:2010zu}: (1) The NR waveform does not have the correct
``memory'' contribution since this depends on the entire history of the system
starting at $t=-\infty$, while the NR simulation covers only the last few
orbits.  (2) The extrapolation to future null infinity does not work as well
for these modes~\cite{SXSCatalog2018}. This could be improved in the future
with Cauchy Characteristic Extraction (CCE)~\cite{Handmer:2015, Handmer:2016,
Winicour2009, Bishop1996}.  (3) The amplitude of these modes is very small
except very close to merger; therefore the early part of the NR waveform where
we compare with the inspiral waveforms is contaminated by numerical noise.

Therefore, when constructing the hybrid waveforms, we set the entire inspiral
waveform to zero for these modes,
\begin{equation}
    \h^{ins}_{\ell, m=0} = 0 \,.
\end{equation}
When computing the cost function~(Eq.~(\ref{Eq:cost_function})), we ignore the
$m=0$ modes.

This means that our hybrid waveforms for these modes are equivalent to the NR
waveforms. In addition, the main contribution for these modes comes from the
region close to merger, which does not correspond to a memory signal, but
instead is due to axisymmetric excitations near merger (cf. bottom panel of
Fig.~\ref{Fig:Hybridization}).

\subsection{Choice of matching region}
\label{Subsec:matching_region}

\begin{figure*}[thb]
\includegraphics[width=\textwidth]{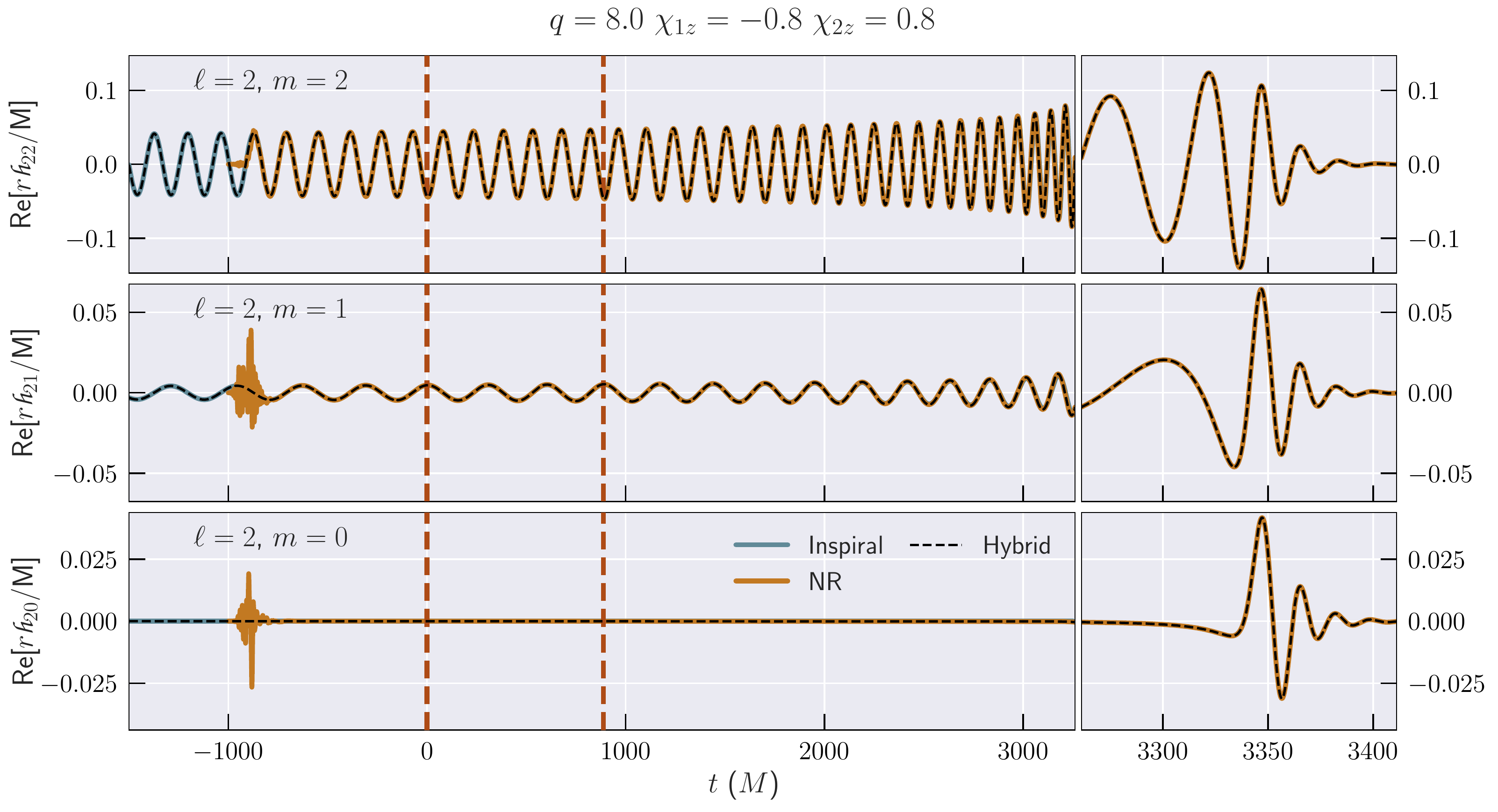}
\caption{
An example hybrid waveform used in this work. We show the $\ell=2$ modes of the
inspiral, NR and hybrid waveforms. The binary parameters are shown on the top
of the plot. The vertical red dashed lines indicate the matching region. Note
that this plot shows the inspiral and NR waveforms after the time and frame
shifts are done.
}
\label{Fig:Hybridization}
\end{figure*}

There are several considerations to take into account when choosing a matching
region $[t_1,t_2]$ for the cost function~(Eq.~(\ref{Eq:cost_function})): (1)
The NR and inspiral waveforms should agree with each other reasonably in this
region; at early times the NR waveform is contaminated by junk radiation while
at late times the inspiral waveform deviates from NR (cf.
Figs.~\ref{Fig:EOB_correction} and \ref{Fig:Hybridization}).  (2) The matching
region should be wide enough that the cost function is meaningful.

Our matching region starts at $1000M$ after the start of the NR waveform; we
find that this is necessary to avoid noise due to junk radiation in some of the
higher order modes. The length of the NR waveforms from the start of the
matching region to the peak of the waveform amplitude varies between $3270M$
and $4227M$. The width of the matching region is then chosen to be equal to the
time taken for 3 orbits of the binary. We use the phase of the (2,2) mode of
the NR waveform to determine this. This choice ensures the width of the
matching region scales appropriately with the NR starting frequency, so that we
get wider matching regions when the NR waveform starts early in the inspiral.

\subsection{Allowed ranges for frame and time shifts}

The allowed range for $\varphi_0$ is [0,$2\pi$].  For nonprecessing binaries
the allowed values for $\psi$ can be restricted by taking into account the
symmetries of the system. We will show that this restriction is a consequence
of the well-known relationship
\begin{gather}
\label{Eq:negm_symmetry}
    \mathpzc{h}_{\ell,-m} = (-1)^{\ell} ~\mathpzc{h}^{*}_{\ell,m} \, ,
\end{gather}
between the $m<0$ modes and the $m>0$ modes for nonprecessing binaries orbiting
in the $x$-$y$ plane~\cite{Boyle:2014}. We compute the shifted waveform
\begin{align}
\mathpzc{h}_{\ell,-m}(t) \, e^{-\mathrm{i} m \varphi_0} \, e^{2 \mathrm{i} \psi} & =
\mathpzc{h}_{\ell,-m}(t; \varphi_0, \psi) \nonumber \\
& =(-1)^{\ell}  (\mathpzc{h}_{\ell,m}(t; \varphi_0, \psi))^{*} \nonumber  \\
& =(-1)^{\ell} e^{-2 \mathrm{i} \psi} e^{-\mathrm{i} m \varphi_0} \mathpzc{h}^{*}_{\ell,m}(t) \nonumber  \\
& = e^{-2 \mathrm{i} \psi} e^{-\mathrm{i} m \varphi_0} \mathpzc{h}_{\ell,-m}(t) \nonumber  \\
\implies e^{2 i \psi} = e^{-2 i \psi}.
    \label{Eq:psi_symm1}
\end{align}
Eq.~(\ref{Eq:psi_symm1}) implies that the only allowed values for $\psi$ are
$0$ and $\pi/2$ \footnote{$\psi=\pi$ is also allowed, but it is degenerate with
$\psi=0$.}. If the inspiral waveform and the NR waveform have the same sign
convention, then $\psi=0$.  Unfortunately, not all NR catalogs and PN-waveform
codes use the same sign convention, so we allow the possibility of $\psi=\pi/2$
to account for this.

To set the allowed range for $t_0$, we begin by computing the orbital frequency
of the inspiral waveform, $\omega^{\ins}$, as half the frequency of the (2,2)
mode. Similarly, we compute the orbital frequency of the NR waveform,
$\omega^\mathrm{NR}$. We first time-align the NR and inspiral waveforms such
that their frequencies match at the start of the matching region. This gives us
a good starting point to vary the time shift.

We also define,
\begin{gather}
\omega^{\ins}_\mathrm{mid} =\omega^\mathrm{NR}(t=t_1) \,, \\
\omega^{\ins}_\mathrm{low} = 0.995\times \omega^{\ins}_\mathrm{mid} \,,
\label{Eq:timeshift_omega_low} \\
\omega^{\ins}_\mathrm{hi} = 1.005\times \omega^{\ins}_\mathrm{mid} \,,
\label{Eq:timeshift_omega_hi}
\end{gather}
where $\omega^\mathrm{NR}(t=t_1)$ is the NR frequency at the start of the
matching region. The allowed range for time shifts $t_0$ is restricted to lie
in the interval [$t^{\ins}_\mathrm{low}-t^{\ins}_\mathrm{mid}$,
$t^{\ins}_\mathrm{hi} - t^{\ins}_\mathrm{mid}$], where $t^{\ins}_\mathrm{low}$,
$t^{\ins}_\mathrm{mid}$ and $t^{\ins}_\mathrm{hi}$ are the times at which
$\omega^{\ins}(t)$ is equal to $\omega^{\ins}_\mathrm{low}$,
$\omega^{\ins}_\mathrm{mid}$ and $\omega^{\ins}_\mathrm{hi}$, respectively. In
other words, the allowed range for $t_0$ is a region near $t_0=0$. $t_0=0$ is
the case when the frequencies of the inspiral and the NR waveforms match at
$t_1$, the start of the matching region. The lower (upper) limit for $t_0$ is
chosen such that the inspiral waveform has a frequency equal to $0.995$
($1.005$) times the NR frequency at $t_1$.

The factors in Eqs.~(\ref{Eq:timeshift_omega_low}) and
(\ref{Eq:timeshift_omega_hi}) are chosen such that the time shift that
minimizes the cost function is always well within the range of allowed time
shifts.  Hence, choosing a wider range (i.e. values of these factors farther
from unity) does not improve the hybridization procedure.  Note also that, like
the width of the matching region in Sec.~\ref{Subsec:matching_region}, setting
the range of time shifts based on the orbital frequency ensures that it scales
appropriately with the start frequency of the NR waveform.

The minimization in Eq.~(\ref{Eq:frame_optimization}) is performed as follows.
We vary the time shift $t_0$ over 500 uniformly spaced values in the above
mentioned time range~\footnote{We find that increasing the number of time
    samples results in no noticeable improvement; the typical values of the
cost function after minimization with 500 samples are $\mathcal{E}\sim10^{-5}$,
and using 1000 samples results in changes only of order
$\Delta\mathcal{E}\lesssim10^{-8}$.}.  For each of these time shifts $t_0$, we
try both allowed values of $\psi \in \{0,\pi/2\}$. For each $t_0$ and $\psi$,
we minimize the cost function over $\varphi_0$ using the Nelder-Mead down-hill
simplex minimization algorithm as implemented in Scipy~\cite{scipy}.
To avoid local minima in the $\varphi_0$ minimization, we perform 10 searches
with different initial guesses, which are sampled from a uniform random
distribution in the range $[0,2\pi]$.

\begin{figure*}[thb]
\makebox[\linewidth][c]{
\includegraphics[width=0.5\textwidth]{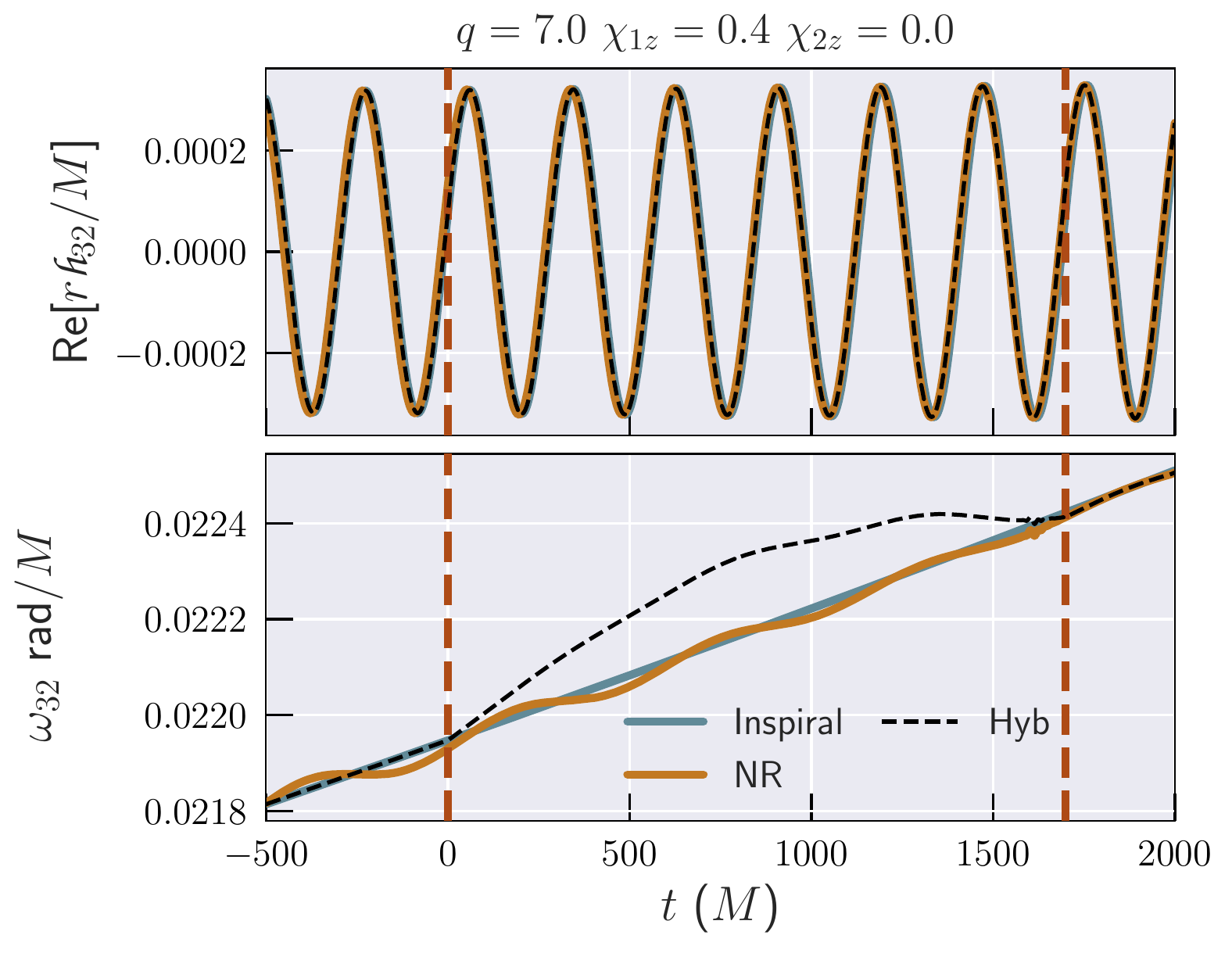}
\includegraphics[width=0.5\textwidth]{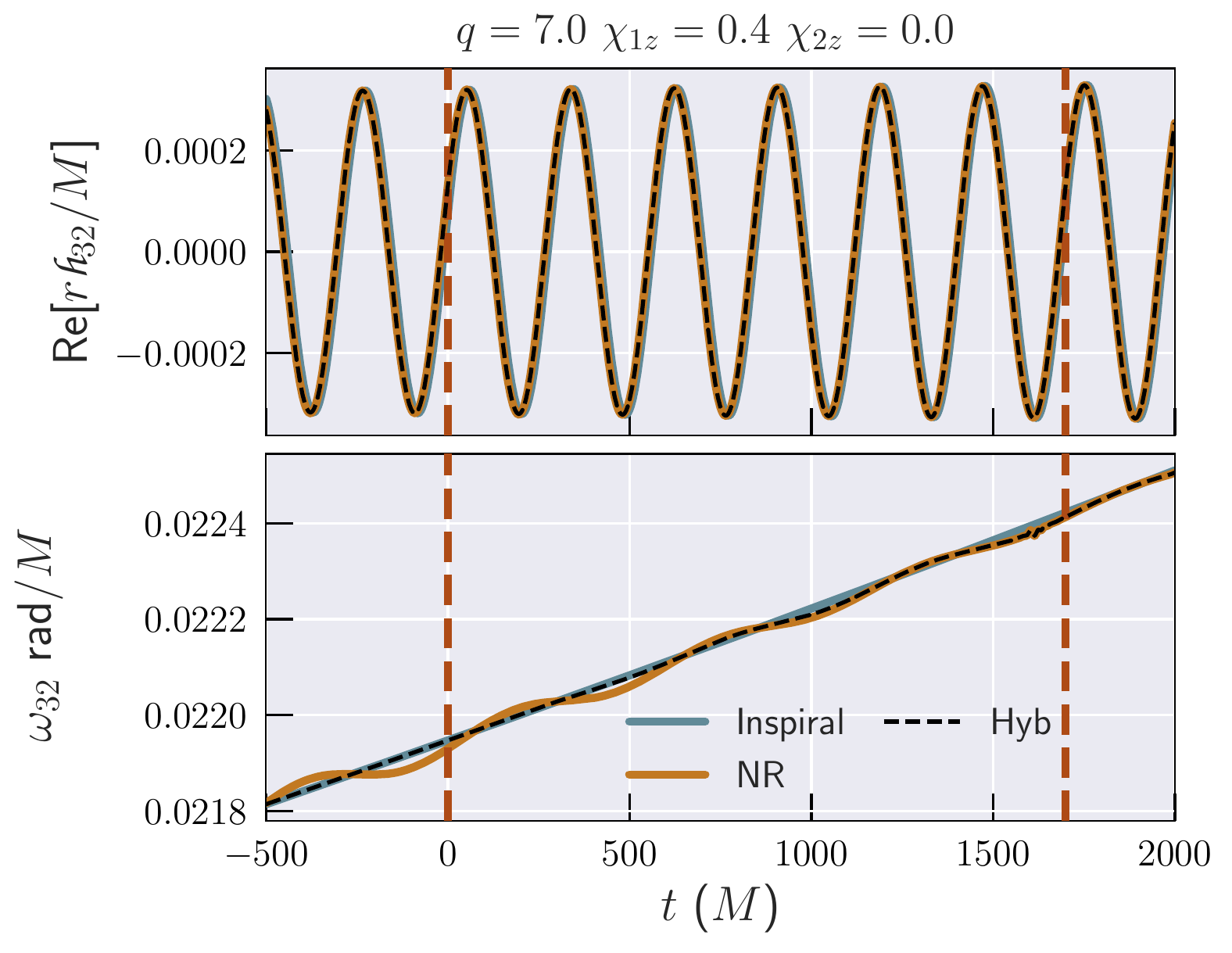}
}
\caption{
{\bf Left}: The real part (top) and frequency (bottom) of the $(3,2)$ mode
using the inertial frame stitching described in
Sec.~\ref{Subsubsec:inertial_stitching}. The binary parameters are shown on the
top of the plot. The vertical red dashed lines indicate the matching region.
Note that this plot shows the inspiral and NR waveforms after the time and
frame shifts are performed.  {\bf Right}: Same, but using the
amplitude-frequency stitching described in
Sec.~\ref{Subsubsec:ampfreq_stitching}. Now we see that the frequency of the
hybrid waveform agrees much better with the NR and inspiral data.
}
\label{Fig:Hyb_inertial_stitching}
\end{figure*}

\subsection{Stitching NR and inspiral waveforms.}
Having obtained the right frame and time shifts between the NR and inspiral
waveforms, the final step is to smoothly stitch the inspiral waveform to the
shifted NR waveform. The stitching is done using a smooth blending function:
\begin{gather}
\tau(t) =
    \begin{cases}
      0, & \text{if}\ t<t_1 \\
      \sin^2\left(\frac{\pi}{2} \frac{t - t_1}{t_2 -t_1}\right),
            & \text{if}\ t_1 \leq t \leq t_2 \\
      1, & \text{if}\ t>t_2\,,
    \end{cases}
\end{gather}
where $t_1$ and $t_2$ take on the same values as those appearing in
Eq.~\eqref{Eq:cost_function}.  Different blending functions have been proposed
in the literature~\cite{Ajith-Babak-Chen-etal:2007b, MacDonald:2011ne,
Santamaria:2010yb, Ajith:2012az}. Our choice is equivalent to the blending function defined
in Ref.~\cite{MacDonald:2011ne}.  We find that our results are not sensitive to
the choice of blending function.

In what follows, for brevity, we drop the hybridization parameters $\varphi_0$,
$\psi$, $t_0$ with the understanding that the models are stitched together
after transforming into hybridization frame,
\begin{gather}
    \hlm^{\ins}(t) \equiv \hlm^{\ins}(t; t_0), \\
    \hlm^{\mathrm{NR}}(t) \equiv \hlm^\mathrm{NR}(t; \varphi_0, \psi).
\end{gather}

Given the shifted waveforms and the blending function, there are still several
ways in which one can stitch the waveforms together.

\subsubsection{Inertial frame stitching}
\label{Subsubsec:inertial_stitching}
One could work with the complex waveform strain and define:
\begin{gather}
\hlm^\mathrm{Hyb} = (1 - \tau(t)) ~\hlm^{\ins}(t)
    + \tau(t)~\hlm^\mathrm{NR}(t) \, .
\end{gather}
With this choice, by construction, the complex strain transitions smoothly from
the inspiral part to the NR part over the matching region.  However, the
transition is more complicated for the frequency, since it involves time
derivatives of the complex argument of the strain; the time derivatives of the
blending function do not behave like a smooth blending function. This is
demonstrated in the left panel of Fig.~\ref{Fig:Hyb_inertial_stitching}: the
inspiral and NR frequencies agree well in the matching region but the frequency
of the hybrid waveform deviates from this.

\subsubsection{Amplitude-Frequency stitching}
\label{Subsubsec:ampfreq_stitching}

To avoid the undesirable artifacts described above, we choose to perform the
inspiral-NR stitching using the amplitude and frequency rather than the
inertial frame strain.

We begin by decomposing the NR and inspiral waveforms into their respective
amplitude and phase:
\begin{align}
\hlm^\mathrm{NR}(t) = \Alm^\mathrm{NR} \mathrm{e}^{- \mathrm{i} \philm^\mathrm{NR} } \,, \quad
\hlm^\mathrm{NR}(t) = \Alm^{\ins} \mathrm{e}^{- \mathrm{i} \philm^{\ins} } \,.
\end{align}
The frequency of each mode
\begin{gather}
\omegalm^\mathrm{NR} = \frac{d\philm^\mathrm{NR}}{dt} \,, \qquad
\omegalm^{\ins} = \frac{d\philm^{\ins}}{dt} \,,
\end{gather}
is then numerically computed from 4th-order finite difference approximations to
the time derivative.  Finally, we stitch the amplitude and frequency of each
mode to get their hybrid versions:
\begin{gather}
\Alm^\mathrm{Hyb} = (1 - \tau(t)) ~\Alm^{\ins}(t) + \tau(t)~\Alm^\mathrm{NR}, \\
\omegalm^\mathrm{Hyb} = (1 - \tau(t)) ~\omegalm^{\ins}(t) \
    + \tau(t)~\omegalm^\mathrm{NR} \,.
\end{gather}

To get the inertial frame strain we first need to integrate the frequency to
get the phase. However, we already know the phase in the region before (only
inspiral) and after (only NR) the matching region. So, we integrate the hybrid
frequency
\begin{gather}
\philm^\mathrm{Hyb-match-region} = \int_{t_1}^{t_2} \omegalm^\mathrm{Hyb} dt \,,
\end{gather}
in the matching region using a 4th-order accurate Runge-Kutta scheme.

Finally, we set the phase of the hybrid waveform to,
\begin{gather}
\philm^\mathrm{Hyb} =
    \begin{cases}
    \philm^{\ins}+ \delta^{1}_{\ell m},
        & \text{if}\ t<t_1 \\
    \philm^\mathrm{Hyb-match-region} + \delta^{2}_{\ell m},
        & \text{if}\ t_1 \leq t \leq t_2 \\
    \philm^\mathrm{NR}, & \text{if}\ t>t_2 \,,
    \end{cases}
\end{gather}
where $\delta^{1}_{\ell m}$ and $\delta^{2}_{\ell m}$ are chosen such that
$\philm^\mathrm{Hyb}$ is continuous at $t_1$ and $t_2$.

Since, by construction, the frequency transitions smoothly from the
inspiral-waveform to NR data, we eliminate the artifact seen in the bottom left
panel of Fig~\ref{Fig:Hyb_inertial_stitching} (dashed line), as demonstrated in
the right panel of Fig.~\ref{Fig:Hyb_inertial_stitching}.

We note that since the $m=0$ modes are purely real/imaginary and nonoscillatory
for nonprecessing systems, they do not have a frequency associated with them,
therefore we use the inertial frame stitching of
Sec.~\ref{Subsubsec:inertial_stitching} for these modes. For these modes the
waveform goes from zero to the NR value over the matching region.

In the hybridized waveform we include the $\ell \leq 4$ and $(5,5)$ modes, but
not the $(4,1)$ or $(4,0)$ modes.  For the $(4,1)$ and $(4,0)$ modes we find
that the inspiral and NR waveforms do not agree very well. This is possibly due
to issues in the extrapolation to future-null infinity~\cite{Boyle-Mroue:2008}
for these modes, and could be resolved in the future with
CCE~\cite{Handmer:2015, Handmer:2016, Winicour2009, Bishop1996}  An example of
the final NR, inspiral and hybrid waveforms is shown in
Fig.~\ref{Fig:Hybridization}.

\section{Building the surrogate model}
\label{Sec:buildSurrogate}

Starting from the 104 NR waveforms described in Sec.~\ref{Sec:TrainingSet} and
Sec.~\ref{Sec:NR_runs}, we construct hybrid waveforms as described in
Sec.~\ref{Sec:Hybridization}. In this section we describe our method to
construct a surrogate model for these hybrid waveforms.

\subsection{Processing the training data}
\label{Subsec:processing_data}

Before building a surrogate model, we process the hybrid waveforms as follows.

\subsubsection{Time shift}

We shift the time arrays of the hybrid waveforms such that the peak of the
total amplitude
\begin{equation}
A_{tot} = \sqrt{\sum_{l,m} |\hlm|^2} \,,
\label{Eq:total_amplitude}
\end{equation}
occurs at $t=0$ for each waveform.

\subsubsection{Frequency and mass ranges of validity}

The length of a hybrid waveform is set by choosing a starting orbital frequency
$\omega_0$, for the inspiral waveform; we use $\omega_0=2 \times 10^{-4} ~
\text{rad}/M$ for all waveforms.  However, for the same starting frequency, the
length in time of the waveform is different for different mass ratios and
spins.  Since we want to construct a time-domain surrogate model, we require a
common time array for all hybrid waveforms. The initial time for the surrogate
is determined by the shortest hybrid waveform in the training data set; this
waveform begins at a time $\sim 5.4 \times 10^8 M$ before the peak.  We
truncate all hybrid waveforms to this initial time value.

The largest starting orbital frequency among the truncated hybrid waveforms is
$\omega_0 = 2.9 \times 10^{-4} ~ \text{rad}/M$, which sets the low frequency
limit of validity of the surrogate.  For LIGO, assuming a starting GW frequency
of $20~\text{Hz}$, the $(2,2)$ mode of the surrogate is valid for total masses
$M \geq 0.9 M_{\odot}$.  The highest spin-weighted spherical harmonic mode we
include in the surrogate model is $(5,5)$, for which the frequency is $5/2$
times that of the $(2,2)$ mode.  Therefore, all modes of the surrogate are
valid for $M \geq 2.25 M_{\odot}$. This coverage of total mass is sufficient to
model all stellar mass binaries of interest for ground based detectors; for an
equal mass binary neutron star system, the total mass is $M \sim 2.7
M_{\odot}$.

\subsubsection{Downsampling and common time samples}

Because the hybrid waveforms are so long, it is not practical to sample the
entire waveform with the same step size we use for the NR waveforms ($0.1 M$).
Fortunately, the early low-frequency portion of each waveform requires sparser
sampling than the later high-frequency portion.  We therefore down-sample the
time arrays of the truncated hybrid waveforms to a common set of time samples.
We choose these samples so that there are 5 points per orbit for the
above-mentioned shortest hybrid waveform in the training data set,
except for $t \geq -1000M$ we choose uniform time samples separated by $0.1M$.
This ensures that we have a denser sampling rate at late times when the
frequency is higher.  We retain times up to $135 M$, which is sufficient to
capture the entire ringdown.

Before downsampling, we first transform the waveform into the co-orbital
frame, defined as:
\begin{gather}
\label{Eq:coorb_frame}
\hlm^C = \hlm ~e^{\mathrm{i} m \phi_\mathrm{orb}}, \\
\label{Eq:AmpPhase_22}
\h_{22} = A_{22} ~e^{-\mathrm{i} \phi_{22}}, \\
\phi_\mathrm{orb} = \frac{\phi_{22}}{2},
\label{Eq:orb_phase}
\end{gather}
where $\hlm$ is the inertial frame hybrid waveform, $\phi_\mathrm{orb}$ is the
orbital phase, and $\phi_{22}$ is the phase of the $(2,2)$ mode. The co-orbital
frame can be thought of as roughly co-rotating with the binary, since we
perform a time-dependent rotation given by the instantaneous orbital phase.
Therefore the waveform is a slowly varying function of time in this frame,
increasing the accuracy of interpolation to the chosen common time samples.
For the $(2,2)$ mode we save the downsampled amplitude $A_{22}$ and
phase $\phi_{22}$, while for all other modes we save $\hlm^C$. We find that
this down-sampling results in interpolation errors $\mathcal{E} \lesssim
10^{-10}$ (defined in Eq.~(\ref{Eq:cost_function})) for all hybrid waveforms.

\subsubsection{Phase alignment}
\label{Subsec:phase_alignment}

After down-sampling to the common temporal grid of the surrogate, we rotate the
waveforms about the z-axis such that the orbital phase $\phi_\mathrm{orb}$ is
zero at $t=-1000 M$.  Note that this by itself would fix the physical rotation
up to a shift of $\pi$. When generating the inspiral waveforms for
hybridization, we align the system such that the heavier BH is on the positive
$x-$axis at the initial frequency; this fixes the $\pi$ ambiguity. Therefore,
after this phase rotation, the heavier BH is on the positive $x-$axis at
$t=-1000 M$ for all waveforms\footnote{Here the BH positions at $t=-1000M$ are
defined from the waveform at future null infinity, using a phase rotation
relative to the early inspiral where the BH positions are well-defined in PN
theory; these positions do not necessarily correspond to the (gauge-dependent)
coordinate BH positions in the NR simulation.}.

\subsection{Decomposing the data}
\label{Subsec:waveform_decomp}

It is much easier to build a model for slowly varying functions of time.
Therefore, rather than work with the inertial
frame strain $\hlm$, which is oscillatory, we work with
simpler ``waveform data pieces'' , as explained below.  We
build a separate surrogate for each waveform data piece. When evaluating the
full surrogate model, we first evaluate the surrogate of each data piece and
then recombine the data pieces to get the inertial frame strain.

A common choice in literature when working with nonprecessing waveforms
has been to decompose the complex strain into an amplitude and phase, each of
which is a slowly varying function of time:
\begin{equation}
\hlm = A_{\ell m} e^{-\mathrm{i} \phi_{\ell m}}.
\label{Eq:AmpPhase}
\end{equation}
However, when $q=1$ and $\chi_{1z}=\chi_{2z}$, the amplitude of odd-$m$ modes
becomes zero due to symmetry. This means that the phase becomes meaningless, so
one has to treat such cases separately.  For example,
Ref~\cite{Blackman:2015pia} used specialized basis functions for the odd-$m$
modes that captured the divergent behavior of the phase in the equal-mass
limit.

To avoid this issue, instead of using the amplitude and phase we use the
real and imaginary parts of the co-orbital frame strain $\hlm^C$,
defined in Eq.~(\ref{Eq:coorb_frame}), for all nonquadrupole modes.  The
co-orbital frame strain is always meaningful: in the special, symmetric case
mentioned above, the co-orbital frame strain for the odd-$m$ modes just goes to
zero, rather than diverge. For the $(2,2)$
mode we use the amplitude\footnote{Note that for the (2,2) mode
$A_{22}=\mathpzc{h}^C_{22}$.} $A_{22}$ and phase $\phi_{22}$.

As mentioned above, our hybrid waveforms are very long, typically containing
$\sim 3 \times 10^4$ orbits.  This presents new challenges that are not present
for pure-NR surrogates. For instance, $\phi_{22}$ sweeps over $\sim 4 \times
10^{5}$ radians for a typical hybrid waveform. We find that the accuracy of the
surrogate model at early times improves if we first subtract a PN-derived
approximation
to the phase, model the phase difference rather
  than $\phi_{22}$, and then add back the PN contribution when evaluating
  the surrogate model. In particular, we use the leading order TaylorT3
approximant~\cite{Damour2001}. For this
approximant, the phase is given
as an analytic, closed-form, function of time.  Therefore, even though TaylorT3
is known to be less accurate than some other approximants~\cite{Buonanno:2009},
its speed makes it ideal for our purpose as we only need it to capture the
general trend. At leading order, the TaylorT3 phase is given by:
\begin{gather}
    \phi^{T3}_{22} = \phi^{T3}_\mathrm{ref} - \frac{2}{\eta ~\theta^5},
    \label{Eq:TaylorT3_phase}
\end{gather}
where $\phi^{T3}_\mathrm{ref}$ is an arbitrary integration constant,
$\theta=[\eta~ (t_\mathrm{ref} - t)/(5 M)]^{-1/8}$, $t_\mathrm{ref}$ is an
arbitrary time offset, and $\eta$ is the symmetric mass ratio. Note that
$\phi^{T3}_{22}$ diverges at $t=t_\mathrm{ref}$. We choose
$t_\mathrm{ref}=1000M$, long after the end of the waveform (recall that the
peak is at $t=0$), to ensure that we are always far away from this divergence.
We choose $\phi^{T3}_\mathrm{ref}$ such that $\phi^{T3}_{22}=0$ at $t=-1000M$;
this is the same time at which we align the hybrid phase in
Sec.~\ref{Subsec:phase_alignment}.

Instead of modeling $\phi_{22}$, we model the residual
\begin{gather}
    \phi^\mathrm{res}_{22} = \phi_{22} - \phi^{T3}_{22} \,,
    \label{Eq:phi22_residual}
\end{gather}
after removing the leading-order contribution $\phi^{T3}_{22}$.  By
construction, $\phi^\mathrm{res}_{22}$ goes to zero at $t=-1000M$.  We find
that after removing the leading order TaylorT3 phase, the scale of
$\phi^\mathrm{res}_{22}$ for a typical hybrid is $\sim 10^3$ radians,
compared to $\sim 4 \times 10^{5}$ radians for $\phi_{22}$. In essence, this
captures almost all of the phase evolution in the early inspiral, simplifying
the problem of modeling the phase to the same as modeling the phase of
late-inspiral NR waveforms.  We stress that the exact form of $\phi^{T3}_{22}$
(or its physical meaning) is not important, as long as it captures the general
trend, since we add the exact same $\phi^{T3}_{22}$ to our model of
$\phi^\mathrm{res}_{22}$ when evaluating the surrogate. In fact, we find that
adding higher order PN terms in Eq.~(\ref{Eq:TaylorT3_phase}) does not improve
the accuracy of the surrogate.

To summarize, we decompose the hybrid waveforms into the following waveform
data pieces, each of which is a smooth, slowly varying function of time:
($A_{22}$, $\phi^\mathrm{res}_{22}$) for the (2,2) mode, and the real and
imaginary parts of $\hlm^C$ for all other modes\footnote{For $m=0$ modes of
nonprecessing systems, $\hlm^C$ is purely real (imaginary) for even (odd)
$\ell$, so we ignore the imaginary (real) part for these modes.}.

\subsection{Building the surrogate}
Once we have the waveform data pieces, we build a surrogate model for each
data piece using the procedure outlined in Refs.~\cite{Blackman:2017dfb,
Field:2013cfa}, which we only briefly describe here. Note that the steps below
are applied independently for each waveform data piece.

\subsubsection{Greedy basis}
We first construct a greedy reduced-basis~\cite{Field:2011mf} such that the
projection errors (cf. Eq.~(5) of Ref.~\cite{Blackman:2017dfb}) for the entire
data set onto this basis are below a given tolerance.  For the basis tolerances
we use $10^{-2}$ radians for the $\phi^\mathrm{res}_{22}$ data piece,
$2\times10^{-5}$ for $A_{22}$, and $8\times10^{-6}$ for all other data pieces.
These are chosen through visual inspection of the basis functions to ensure
they are not noisy, and based on the expected truncation error of the NR
waveforms. For instance, we expect the error in phase to be about $10^{-2}$
radians.

The greedy procedure is initialized with a single basis function as
described in Ref.~\cite{Blackman:2017dfb}. Then at each step in
the greedy procedure, the waveform with the highest projection error onto the
current basis is added to the basis. Previous work has shown that the
resulting greedy reduced-basis is robust to different choices
of initialization~\cite{Herrmann:2012if}. When computing the basis projection
errors, we only include data up to $50 M$ after the peak. We find that this
helps avoid noisy basis functions. This is particularly important for the phase
data piece as this becomes meaningless at late times, when the waveform
amplitude becomes very small.

\subsubsection{Empirical interpolation}
Next, using a different greedy procedure, we construct an empirical
interpolant~\cite{Barrault:2004, Maday:2009, Hesthaven:2014} in time. This
picks out the most representative time nodes, where the number of time nodes is
the same as the number of greedy basis functions. We require that the start of
the waveform always be included as a time node for all data pieces. This is a
useful modeling choice because the magnitude of the waveform data pieces in the
very early inspiral can be smaller than the basis tolerances mentioned above.
By requiring the first index to be an empirical time node, we enforce an anchor
point that ensures the waveform data piece has the right magnitude at the start
of the waveform. Furthermore, we do not allow any empirical time nodes at times
$>50 M$, since we expect this part to be dominated by noise (especially for the
phase data piece).

\begin{figure*}[thb]
\mbox{
\includegraphics[width=0.475\textwidth]{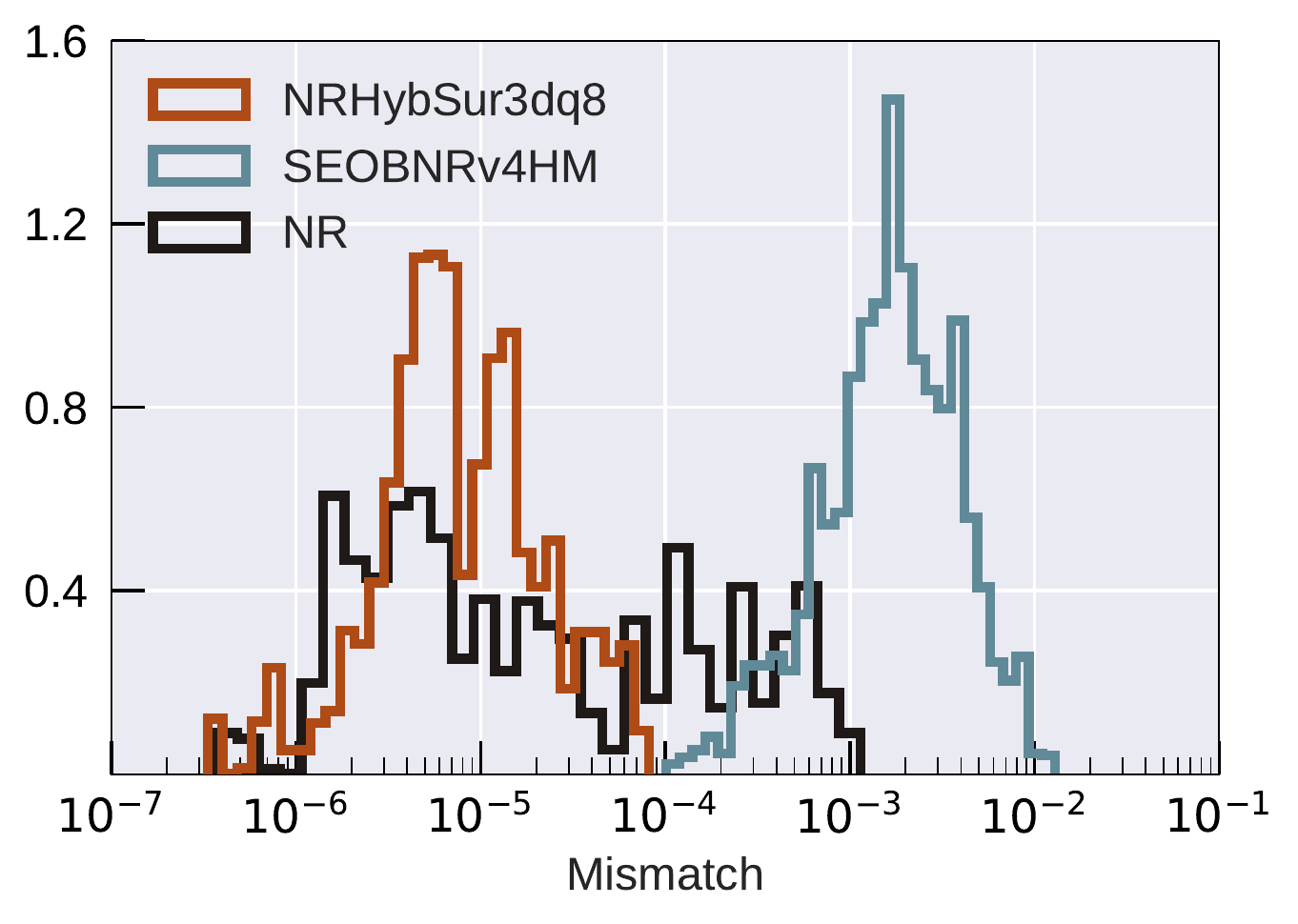}
\includegraphics[width=0.5\textwidth]{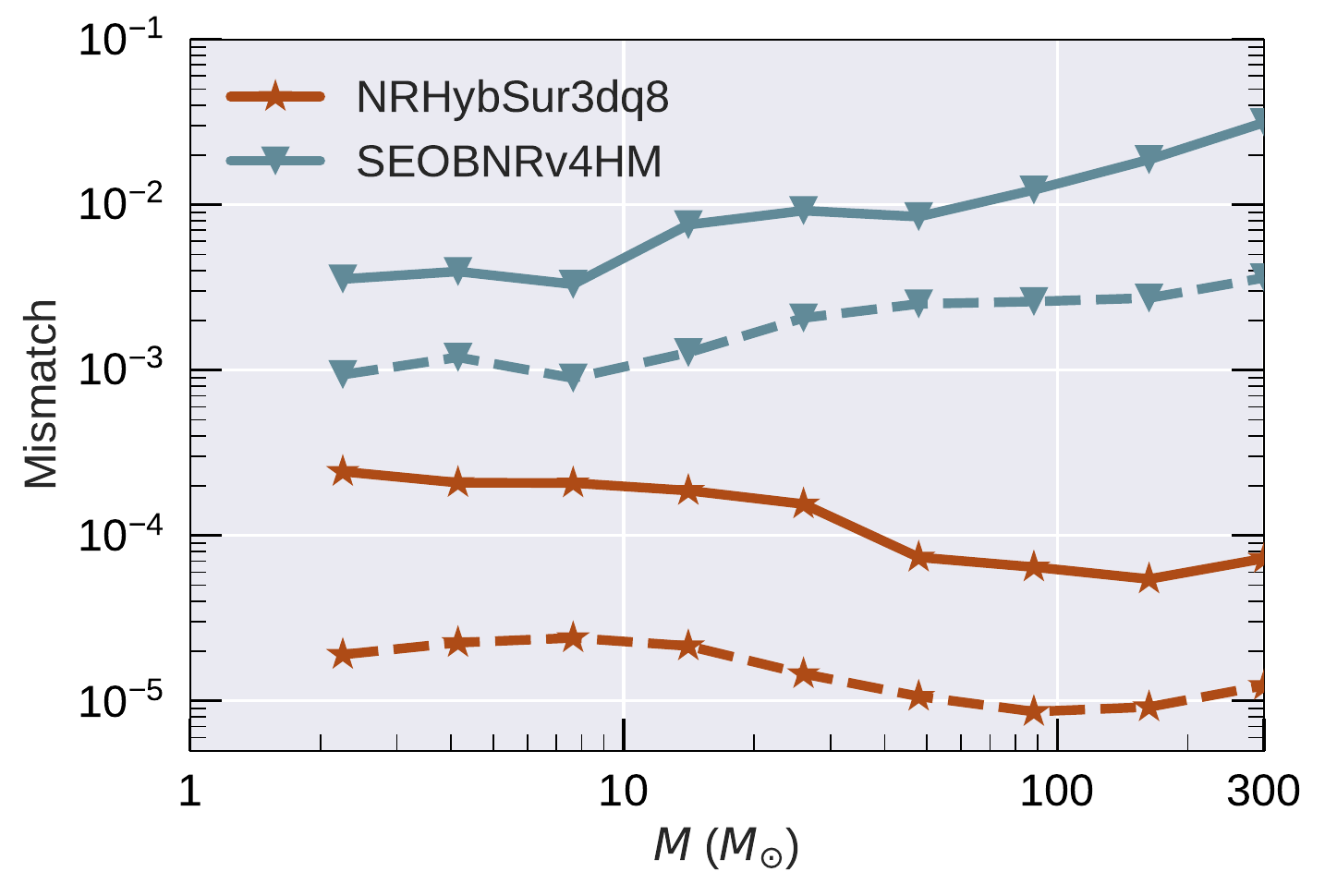}
}
\caption{
Errors in NRHybSur3dq8 and SEOBNRv4HM when compared against hybrid waveforms.
For NRHybSur3dq8, we show out-of-sample errors. Mismatches are computed at
several points in the sky of the source frame using all available modes in each
waveform: For the hybrid waveforms and NRHybSur3dq8, that is $\ell\leq4$ and
$(5,5)$, but not $(4,1)$ or $(4,0)$.  For SEOBNRv4HM that is
$(2,2),(2,1),(3,3),(4,4)$, and $(5,5)$.  {\bf Left}: Mismatches computed using
a flat noise curve, but including only the late inspiral part of the waveforms,
starting at $-3500M$ before the peak. Therefore, we are essentially comparing
only to the NR part of the hybrid waveforms. For comparison, we also show the
NR resolution error, obtained by comparing the two highest available
resolutions. The histograms are normalized such that the area under each curve
is $1$ when integrated over $\log_{10}$(Mismatch). {\bf Right}: Mismatches as a
function of total mass, computed using the Advanced LIGO design sensitivity
noise curve. Here we compare against the full hybrid waveforms. The solid
(dashed) lines show the 95th percentile (median) mismatch values
over points on the sky as well as different hybrid waveforms.
}
\label{Fig:ligo_mismatches}
\end{figure*}

\subsubsection{Parametric fits}
Finally, for each time node, we construct a fit across the parameter space.
The fits are done using the Gaussian process regression (GPR) fitting method
described in the supplemental material of Ref.~\cite{Varma:2018aht}. Following
Ref.~\cite{Varma:2018aht}, we parameterize our fits using $\log(q)$,
$\hat{\chi}$, and $\chi_a$. Here $\hat{\chi}$ is the spin parameter entering
the GW phase at leading order \cite{Khan:2015jqa, Ajith:2011ec,
CutlerFlanagan1994, Poisson:1995ef} in the PN expansion,
\begin{gather}
\chieff = \frac{q~\chi_{1z} + \chi_{2z}}{1+q}\,,\\
\hat{\chi} = \frac{\chieff - 38\eta(\chi_{1z} + \chi_{2z})/113}
    {1-{76\eta}/{113}}\,,
\end{gather}
and $\chi_a$ is the ``anti-symmetric spin'',
\begin{equation}
    \chi_a = \tfrac{1}{2}(\chi_{1z} - \chi_{2z})\,.
\label{eq:anti_symm_spin}
\end{equation}
The fit accuracy, and as a result the accuracy of the surrogate model, improves
noticeably when using $\log(q)$, compared to $q$ or $\eta$.

\begin{figure*}[thb]
\includegraphics[width=\textwidth]{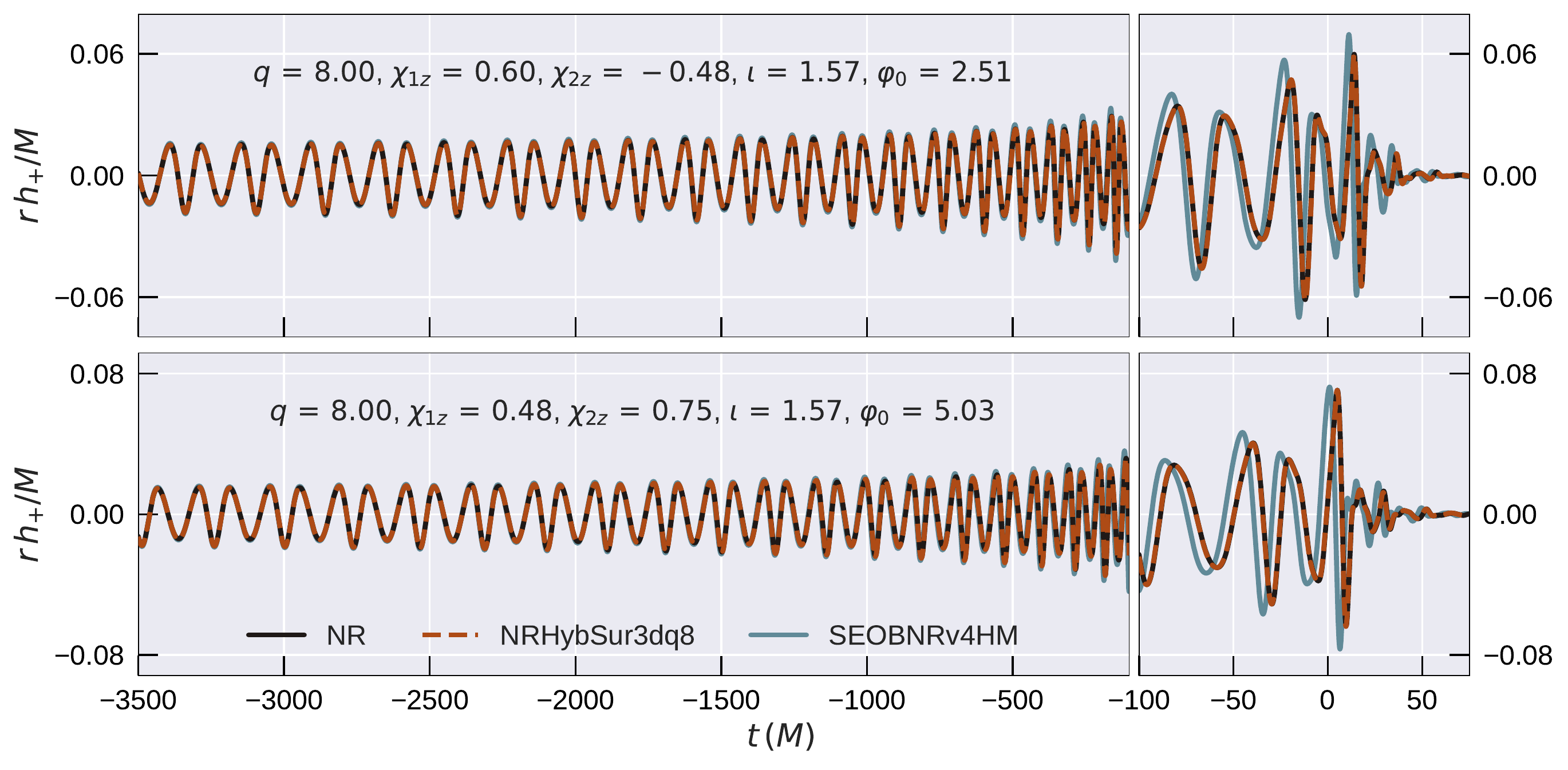}
\caption{The plus polarization of the waveforms for the cases that result in
the largest mismatch for NRHybSur3dq8 (top) and SEOBNRv4HM (bottom) in the left
panel of Fig.~\ref{Fig:ligo_mismatches}. We also show the corresponding hybrid
waveforms (labeled as NR because only the late part is shown). Each waveform is
projected using all available modes for that model, along the direction which
results in the largest mismatch for NRHybSur3dq8 (SEOBNRv4HM) in the top
(bottom) panel.  Note that NRHybSur3dq8 is evaluated using trial surrogates
that are not trained using these cases. The binary parameters and the direction
in the source frame are indicated in the inset text. All waveforms are time
shifted such that the peak of the total waveform amplitude occurs at $t=0$
(using all available modes, according to Eq.~(\ref{Eq:total_amplitude})).  Then
the waveform modes are rotated about the $z-$axis such that the orbital phase
is zero at $t=-3500M$.
}
\label{Fig:worst_case}
\end{figure*}

\subsection{Evaluating the surrogate}

When evaluating the surrogate waveform, we first evaluate each surrogate
waveform data piece.  Next, we compute the phase of the $(2,2)$ mode,
\begin{gather}
    \phi^\mathrm{S}_{22} \equiv \phi^\mathrm{res,S}_{22} + \phi^{T3}_{22},
\end{gather}
where
$\phi^\mathrm{res,S}_{22} \approx \phi^\mathrm{res}_{22}$ is the surrogate
model for $\phi^\mathrm{res}_{22}$ and $\phi^{T3}_{22}$
is
given in Eq.~(\ref{Eq:TaylorT3_phase}). If the waveform is required at a
uniform sampling rate, we interpolate each waveform data piece from the sparse
time samples used to construct the model to the required time samples, using a
cubic-spline interpolation scheme. Finally, we use
Eqs.~(\ref{Eq:coorb_frame}), (\ref{Eq:AmpPhase_22}), and (\ref{Eq:orb_phase})
to reconstruct the surrogate prediction for the inertial frame strain.

\section{Results}
\label{Sec:Results}

In order to estimate the difference between two waveforms, $\h_1$ and $\h_2$,
we use the mismatch, defined in Eq.~(\ref{Eq:time_dom_mismatch}), but in this
section instead of Eq.~(\ref{Eq:time_dom_mismatch_2}) we use the
frequency-domain inner product
\begin{gather}
\left<\h_1, \h_2\right> = 4 \mathrm{Re}
    \int_{f_{\mathrm{min}}}^{f_{\mathrm{max}}}
    \frac{\tilde{\h}_1 (f) \tilde{\h}_2^* (f) }{S_n (f)} df,
\label{Eq:freq_domain_Mismatch}
\end{gather}
where $\tilde{\h}(f)$ indicates the Fourier transform of the complex strain
$\h(t)$, $*$ indicates a complex conjugation, $\mathrm{Re}$ indicates the real
part, and $S_n(f)$ is the one-sided power spectral density of a GW detector. We
taper the time domain waveform using a Planck window~\cite{McKechan:2010kp},
and then zero-pad to the nearest power of two.  We further zero-pad the
waveform to increase the length by a factor of eight before performing the
Fourier transform. The tapering at the start of the waveform is done over $1.5$
cycles of the $(2,2)$ mode. The tapering at the end is done over the last
$20M$. Note that our model contains times up to $135M$ after the peak of the
waveform amplitude, and the signal has essentially died down by the last $20M$.

We compute mismatches following the procedure described in Appendix D of
Ref.~\cite{Blackman:2017dfb}: the mismatches are optimized over shifts in time,
polarization angle, and initial orbital phase.  Both plus and cross
polarizations are treated on an equal footing by using a two-detector setup
where one detector sees only the plus and the other only the cross
polarization. We compute the mismatches at 37 points uniformly distributed on
the sky in the source frame, and we use all available modes of a given waveform
model.

When computing flat noise mismatches ($S_n=1$), we take $f_{\mathrm{min}}$ to
be the frequency of the $(2,2)$ mode at the end of the initial tapering window,
and $f_{\mathrm{max}} = 5 f^{peak}_{22}$, where $f^{peak}_{22}$ is the
frequency of the $(2,2)$ mode at its peak. This choice of $f_{\mathrm{max}}$
ensures that we capture the peak frequencies of all modes considered in this
work, including the $(5,5)$ mode, whose frequency has the highest multiple of
the $(2,2)$ mode frequency of all the modes we model.  We also compute
mismatches with the Advanced-LIGO design sensitivity Zero-Detuned-HighP noise
curve~\cite{LIGO-aLIGODesign-NoiseCurves} with $f_{\mathrm{min}}=20 \text{Hz}$
and $f_{\mathrm{max}}=2000 \text{Hz}$.

\subsection{Surrogate errors}
\label{Subsec:surr_errs}

We evaluate the accuracy of our new surrogate model, NRHybSur3dq8, by computing
mismatches against hybrid waveforms. For this, we compute ``out-of-sample''
errors as follows. We first randomly divide the 104 training waveforms into
groups of $\roughly 5$ waveforms each.  For each group, we build a trial
surrogate using the remaining $\roughly 99$ training waveforms and test against
these five validation ones.  We also compute the mismatch between an existing
higher-mode waveform model, SEOBNRv4HM~\cite{Cotesta:2018fcv}, and the hybrid
waveforms.

Figure~\ref{Fig:ligo_mismatches} summarizes mismatches of both NRHybSur3dq8 and
SEOBNRv4HM versus the hybrid waveforms.  We use all available modes for each
waveform model. In the left panel we show mismatches computed using a flat
noise curve over the NR part of the hybrid waveforms (to do this, we truncate
the waveforms and begin tapering at $t=-3500M$).  We see that the mismatches
for NRHybSur3dq8 are about two orders of magnitude lower than that of
SEOBNRv4HM. We compare this with the truncation error in the NR waveforms
themselves, by computing the mismatch between the two highest available
resolutions of each NR waveform.  The errors in the surrogate model are well
within the truncation error of the NR simulations. Note that NR error
estimated in this manner is a conservative estimate; if we treat the high
resolution simulation as the fiducial case, the NR curve in
Fig.~\ref{Fig:ligo_mismatches} can be thought of as the error in the
lower-resolution simulation.
This explains why the errors in the surrogate are smaller
than the NR errors. We suspect that the error of the high resolution
simulations is close to the surrogate model's error.

The right panel of Fig.~\ref{Fig:ligo_mismatches} shows mismatches computed
using the Advanced LIGO design sensitivity noise curve. The mismatches are now
dependent on the total mass of the system, so we show mismatches for masses
starting at the lower limit of the range of validity of the surrogate:
$M\geq2.25 M_{\odot}$. 95th percentile mismatches for NRHybSur3dq8,
are always below $\roughly3\times10^{-4}$ in the mass range $2.25 M_{\odot}
\leq M \leq 300 M_{\odot}$. At high masses ($M\gtrsim40M_{\odot}$), where
the merger and ringdown are more prominent, our model is more accurate than
SEOBNRv4HM by roughly two orders of magnitude, in agreement with the left panel
of Fig.~\ref{Fig:ligo_mismatches}.

For high masses only the last few orbits of the hybrid waveforms are in the
LIGO band, and the hybrid waveforms are effectively the same as the NR
waveforms.  For low masses, the errors in the right panel of
Fig.~\ref{Fig:ligo_mismatches} quantify how well different models reproduce the
hybrid waveforms. However, this comparison cannot account for the errors in the
hybridization procedure itself. We provide some evidence for the fidelity of
the hybrid waveforms in Sec.~\ref{Subsec:longNRTests}, by comparing against
some long NR waveforms.

Fig.~\ref{Fig:worst_case} shows NRHybSur3dq8 and SEOBNRv4HM waveforms
for the cases leading to the largest errors in the left panel of
Fig.~\ref{Fig:ligo_mismatches}. The surrogate shows very good agreement with
the NR waveform, even for its worst case. SEOBNRv4HM shows a noticeably larger
deviation that cannot all be accounted for with a time and/or phase shift. Note
that we align the time and orbital phase of the waveforms in
Fig.~\ref{Fig:worst_case}.

We note that the main improvement over SEOBNRv4HM is not due to the inclusion
of more modes. We find that the agreement between SEOBNRv4HM and the NR/hybrid
waveforms in Figs.~\ref{Fig:ligo_mismatches} and \ref{Fig:worst_case} improves
only marginally when restricting the NR/hybrid waveforms to the same set of
modes as SEOBNRv4HM.

\subsection{Hybridization errors}
\label{Subsec:longNRTests}

The errors described in Sec.~\ref{Subsec:surr_errs} are computed by comparing
the surrogate against hybrid waveforms, hence they do not include the errors in
the hybridization procedure or the errors from EOB-corrected-PN waveforms (cf.
Sec.~\ref{Subsec:EOB_correction}) we use for the early inspiral.  To estimate
these errors, we compare the surrogate against a few very long NR
simulations~\footnote{Note that for these long NR simulations, the outer
    boundary location is chosen based on the length of the
simulations~\cite{SXSCatalog2018} so as to avoid unphysical center-of-mass
accelerations seen in earlier long-duration runs~\cite{Szilagyi:2015rwa}.}.  We
perform five new simulations that are $\sim10^5 M$ long and two that are
$\sim3\times10^4 M$ long. These have been assigned the identifiers SXS:BBH:1412
- SXS:BBH:1418, and will be made publicly available in the upcoming update of
the SXS public catalog~\cite{SXSCatalog}. In addition, we use two simulations
of length $\sim3\times10^4 M$ from Ref.~\cite{Kumar:2015tha}.  These nine
simulations are represented as square markers in Fig.~\ref{Fig:NRParams}, and
have not been used in training the surrogate. The surrogate was trained against
hybrid waveforms whose NR duration varied between $3270M$ and $4227M$.
Therefore, comparing against long NR waveforms, which include the early
inspiral, is a good way to estimate the hybridization error.

\begin{figure}[thb]
\includegraphics[width=0.5\textwidth]{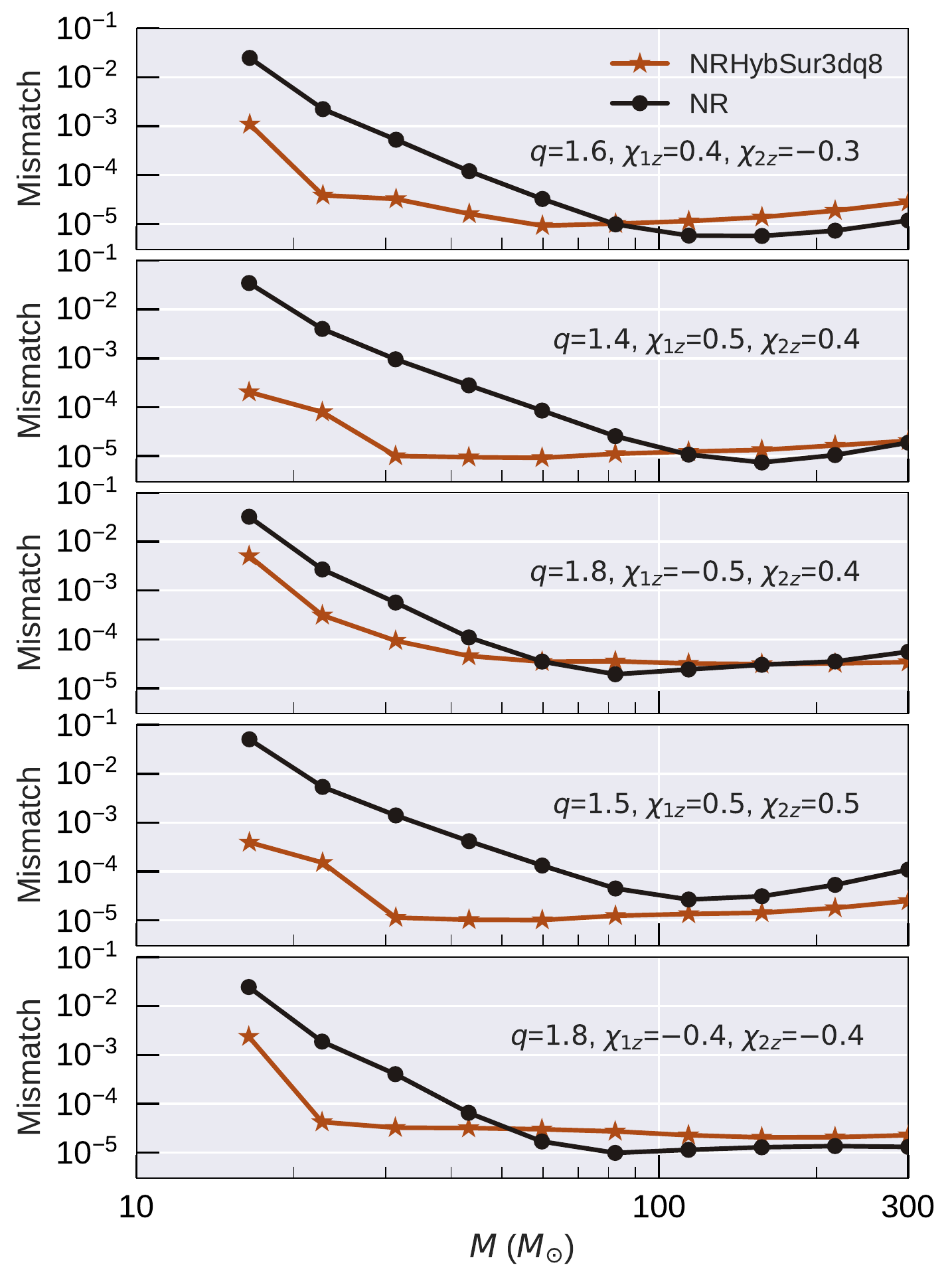}
\caption{Comparisons between the NRHybSur3dq8 surrogate model and a few NR
waveforms of $\roughly 10^5 M$ in duration. We also show the NR resolution
error. 95th percentile mismatches (over points in the sky) are
shown as a function of total mass. The inset text indicates the mass ratio and
component spins. Mismatches are computed using the Advanced LIGO design
sensitivity noise curve. To best assess the error introduced by the
hybridization procedure we use the same set of modes for the NR waveforms as
the surrogate. At low masses, the hybridization errors (red circles)
become less reliable measures of accuracy due to the large NR resolution error
(black circles) itself. Fig.~\ref{Fig:mismatches_hybtest} describes a
refined comparison to improve the assessment at low masses.
}
\label{Fig:ligo_mismatches_hybtest}
\end{figure}

\begin{figure*}[thb]
\includegraphics[width=\textwidth]{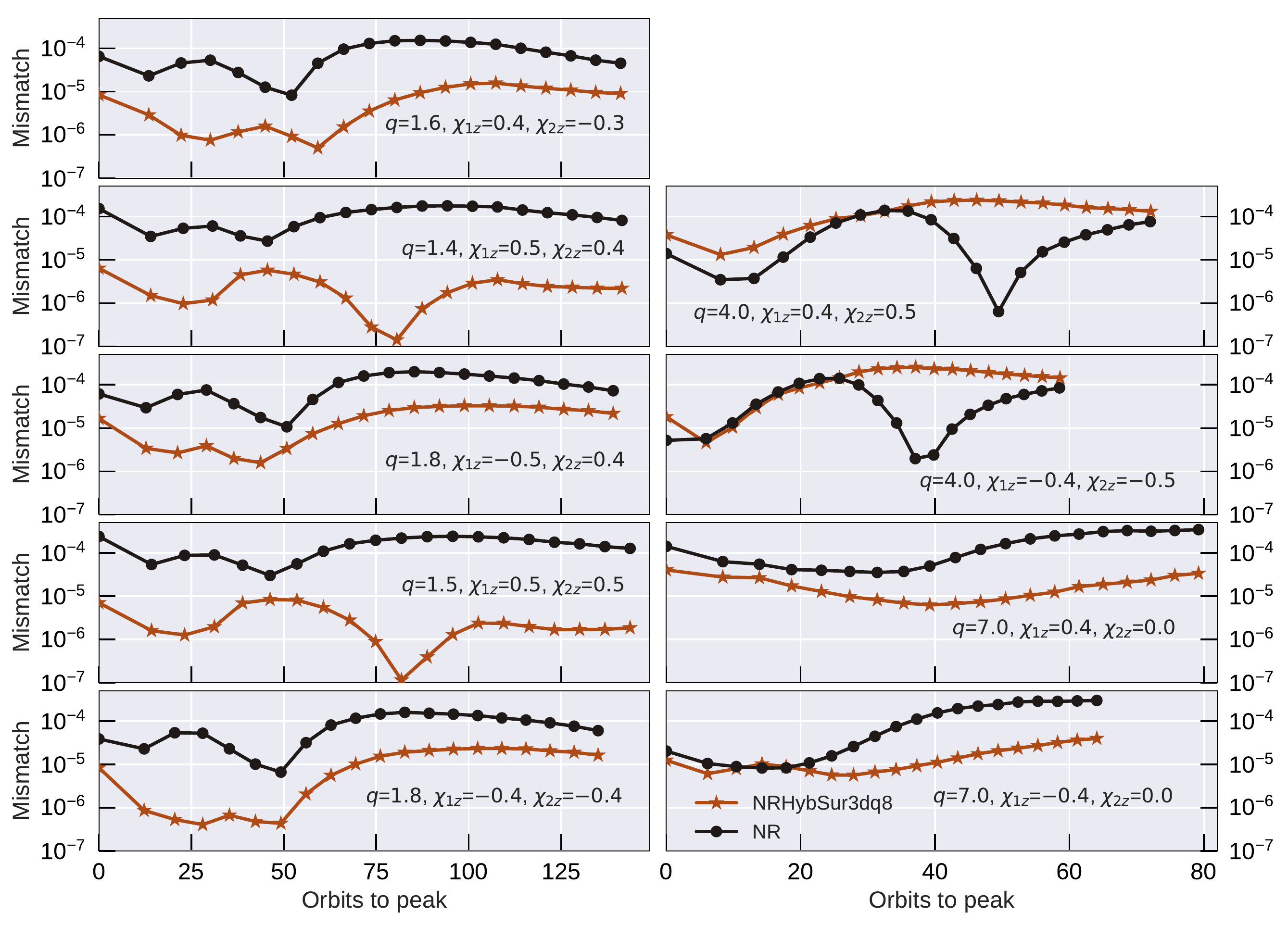}
\caption{Errors in the NRHybSur3dq8 surrogate model against long NR waveforms,
but only looking at segments of length $\Delta t = 5\times10^3 M$ individually.
Each point represents one segment that ends at a specified number of orbits
before the waveform peak, as plotted on the horizontal axis; Therefore, going
from left to right in the figure, we plot segments that start earlier in the
inspiral.  We also show the NR resolution error in the same segments. The inset
text indicates the mass ratio and component spins.  We show
95th percentile mismatches (over points in the sky), computed using
a flat noise curve. We use the same set of modes for the NR waveforms as the
surrogate.  We find that, in general, the surrogate error is lower than or
comparable to the NR resolution error throughout the inspiral.
}
\label{Fig:mismatches_hybtest}
\end{figure*}

We begin by repeating the mismatch computation from the right panel of
Fig.~\ref{Fig:ligo_mismatches}, using the $10^5 M$ long NR waveforms.  This is
shown in Fig.~\ref{Fig:ligo_mismatches_hybtest}. We also show the errors in the
NR simulations, estimated by comparing the two highest available NR
resolutions. We find that the mismatches between the surrogate and the long NR
waveforms for $M>30 M_{\odot}$ are below $10^{-4}$, in agreement with
Fig.~\ref{Fig:ligo_mismatches}.  For lower masses, the mismatches quickly
increase and can be as high as $\sim10^{-2}$. However, this increase in
mismatch is accompanied by an increase in the error of the NR waveforms.  This
is expected, since for very long NR waveforms the accumulated phase error is a
dominant source of numerical error, which becomes increasingly relevant for low
mass systems as more of the waveform moves in-band.  Therefore, in
Fig.~\ref{Fig:ligo_mismatches_hybtest}, at low masses, the comparison between
the surrogate and NR waveforms is largely dominated by the numerical resolution
error of the long NR waveforms themselves.

We find that a better test of the hybridization procedure, one that is less
sensitive to NR phase accumulation errors, is to compare against different
segments of the NR waveform.  Since the phase errors accumulate over a large
number of cycles, by looking at smaller segments we ensure that this
contribution is not the dominant error.  To be precise, we compare the
surrogate and the NR data, using segments of length $\Delta t = 5\times10^{3}
M$ ending at a particular number of orbits before the peak of the waveform. For
each segment we compute mismatches at several points in the sky using a flat
noise curve. By varying the number of orbits to the peak, we can cover the
entire NR waveform including the early inspiral region where the surrogate
depends on the hybridization procedure.  These errors are shown in
Fig.~\ref{Fig:mismatches_hybtest}. We find that in each segment, the mismatch
between the surrogate and the NR data is, in general, lower or comparable to
the NR resolution error. Therefore, the surrogate reproduces the NR data
accurately in the early inspiral and the hybridization errors are smaller than
or comparable to the NR resolution error for these cases. We note that the
surrogate errors in Fig.~\ref{Fig:mismatches_hybtest} depend on the length of
the segment considered and are only meaningful when compared to the NR errors
in the same segment.

Unfortunately, long NR simulations such as these are not available at
regions of the parameter space where both mass ratio and spin magnitudes are
large. These are the cases where PN is expected to perform poorly, so we expect
larger hybridization errors for these cases.

\subsection{Extrapolation outside the training range}
\label{Subsec:extrapTests}

\begin{figure}[bht]
\includegraphics[width=0.5\textwidth]{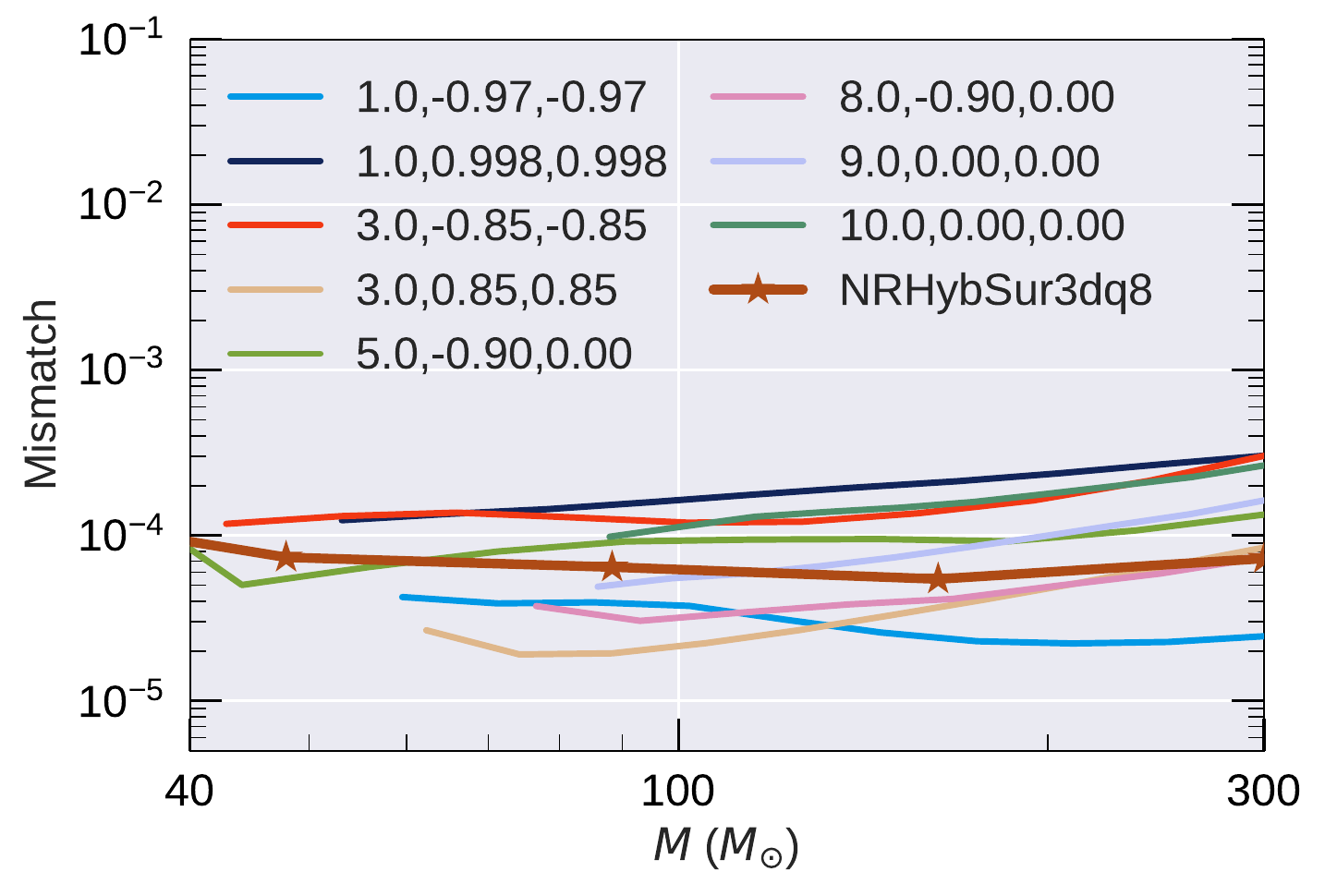}
\caption{Errors in NRHybSur3dq8 when evaluated outside its training range.
95th percentile mismatches (over points in the sky) are shown as a
function of total mass for different extrapolated cases. These are computed
using the Advanced LIGO design sensitivity noise curve. To best assess the
error introduced by the extrapolation, we use the same set of modes for the NR
waveforms as the surrogate.  The labels indicate the mass ratio and component
spins ($q,\chi_{1z},\chi_{2z}$). For comparison we reproduce the
95th percentile mismatches for NRHybSur3dq8 within its training
range from the right panel of Fig.~\ref{Fig:ligo_mismatches}.
}
\label{Fig:extrapNR_mismatches}
\end{figure}

We now investigate the efficacy of NRHybSur3dq8 to extrapolate beyond its
training parameter range by comparing against SpEC NR
simulations~\cite{SXSCatalog, Kumar:2015tha, Chu:2015kft, Scheel2014,
Chatziioannou:2018wqx} at larger mass ratios ($8 < q \le 10$) and/or larger
spin magnitudes ($|\chi_{1z}| > 0.8$ or $|\chi_{12}|> 0.8$).  These NR
simulations are represented as triangle markers in Fig.~\ref{Fig:NRParams}.

Fig.~\ref{Fig:extrapNR_mismatches} shows mismatches for NRHybSur3dq8 when
compared against these simulations. We find that surrogate extrapolates
remarkably well, with the mismatch always $\lesssim4\times10^{-4}$ for all
cases, which include mass ratios up to $q=10$ and spin magnitudes up to
$|\chi|=0.998$. However, the extrapolation errors can be about half an order of
magnitude larger than errors within the training range.  Note that NR
simulations with both high mass ratios and high spin magnitudes are not
currently available, and the ones used here represent the most extreme cases
found in the SXS Catalog. We do not hybridize these simulations before
comparing to NRHybSur3dq8 because several of them are too short. In
Fig.~\ref{Fig:extrapNR_mismatches}, the minimum mass for each case is chosen to
be the lowest mass at which all used modes of the NR simulation lie fully in
the LIGO band with a low frequency cut-off of $20 ~\text{Hz}$.

At much higher mass ratios than those tested here, such as $q=15$, we find that
the waveforms generated by the surrogate can have ``glitches" in the time
series. Therefore, we recommend the surrogate be used for $q\leq10$ and
$|\chi_{1z}|,|\chi_{2z}|\leq1$.
However, we advise caution with any
extrapolation in general.

\subsection{Mode mixing}

\begin{figure}[thb]
\includegraphics[width=0.5\textwidth]{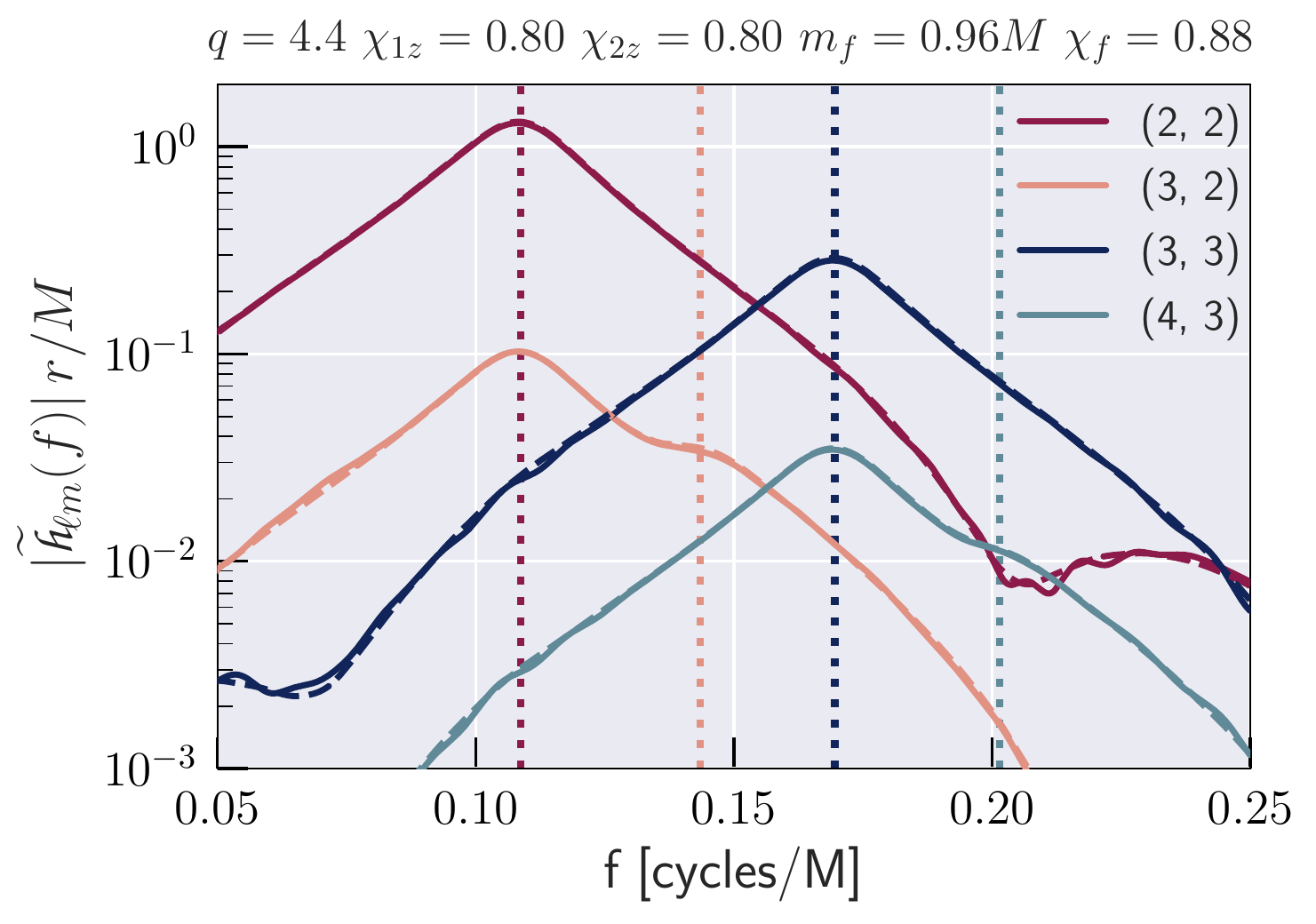}
\caption{Mode mixing between spherical harmonic modes is clearly seen in the
ringdown signal of the NR waveform and is accurately reproduced by the
surrogate. The absolute values of the Fourier transform of different spherical
harmonic modes are shown as solid (dashed) curves for the surrogate (NR).
The dotted vertical lines indicate the frequencies of the fundamental QNM
overtone of these modes. The component parameters as well as the remnant mass
and spin are shown in the text above the figure.
}
\label{Fig:mode_mixing}
\end{figure}

Numerical relativity waveforms are extracted as spin-weighted spherical
harmonic modes~\cite{NewmanPenrose1966, Goldberg1967}. However, in the ringdown
regime, the natural basis to use is the spin-weighted \emph{spheroidal}
harmonic basis~\cite{Teukolsky, Teukolsky:1972my}. A spherical harmonic mode
$\hlm$ can be written as a linear combination of all spheroidal harmonic modes
$\hlm^S$ with the same $m$ index~\cite{Berti:2014fga}. Therefore, during the
ringdown, we expect leakage of power between different spherical harmonic
modes with the same $m$. This is referred to as mode mixing.

Since the surrogate accurately reproduces the spherical harmonic modes from the
NR simulations, it also captures this mode mixing. We demonstrate this for an
example case in Fig.~\ref{Fig:mode_mixing}. Here we compute the Fourier
transform of different spherical harmonic modes in the ringdown stage of the
waveform. Before computing the Fourier transform, we first drop all data before
$t=20M$, where $t=0$ corresponds to the peak of the waveform amplitude (cf.
Eq.~(\ref{Eq:total_amplitude})). Then, we taper the data between $t=20M$ and
$t=40M$, as well as the last $10M$ of the time series, using a Planck
window~\cite{McKechan:2010kp}.  The tapering width at the start is chosen such
that the remaining signal is dominated by the fundamental quasi-normal mode
(QNM) overtone.  Fig.~\ref{Fig:mode_mixing} shows the absolute value of these
Fourier transforms for different modes, for both the surrogate and the NR
waveform.  In addition, we show the frequency of the fundamental QNM overtone
for each mode~\cite{BertiWebsite}.

Note that the $(2,2)$ mode and the $(3,2)$ mode have the same $m$ index, the
condition required for mode mixing. We see that the peak of the $(2,2)$ mode
agrees with the QNM frequency as expected. For the $(3,2)$ mode however, while
there are features of a peak at the expected QNM frequency, there is a much
larger peak at the frequency of the $(2,2)$ mode. This is because some of the
power of the stronger $(2,2)$ mode has leaked into the $(3,2)$ mode due to
mode mixing.  Mode mixing can also be seen for the $(3,3)$ and $(4,3)$ modes,
which also have the same $m$ index. Fig.~\ref{Fig:mode_mixing} shows that not
only does the surrogate agree with NR in the ringdown, it also reproduces the
mode mixing present in the NR data.

\subsection{Evaluation cost}
\label{Subsec:evaluation_cost}

\begin{figure}[thb]
\includegraphics[width=0.5\textwidth]{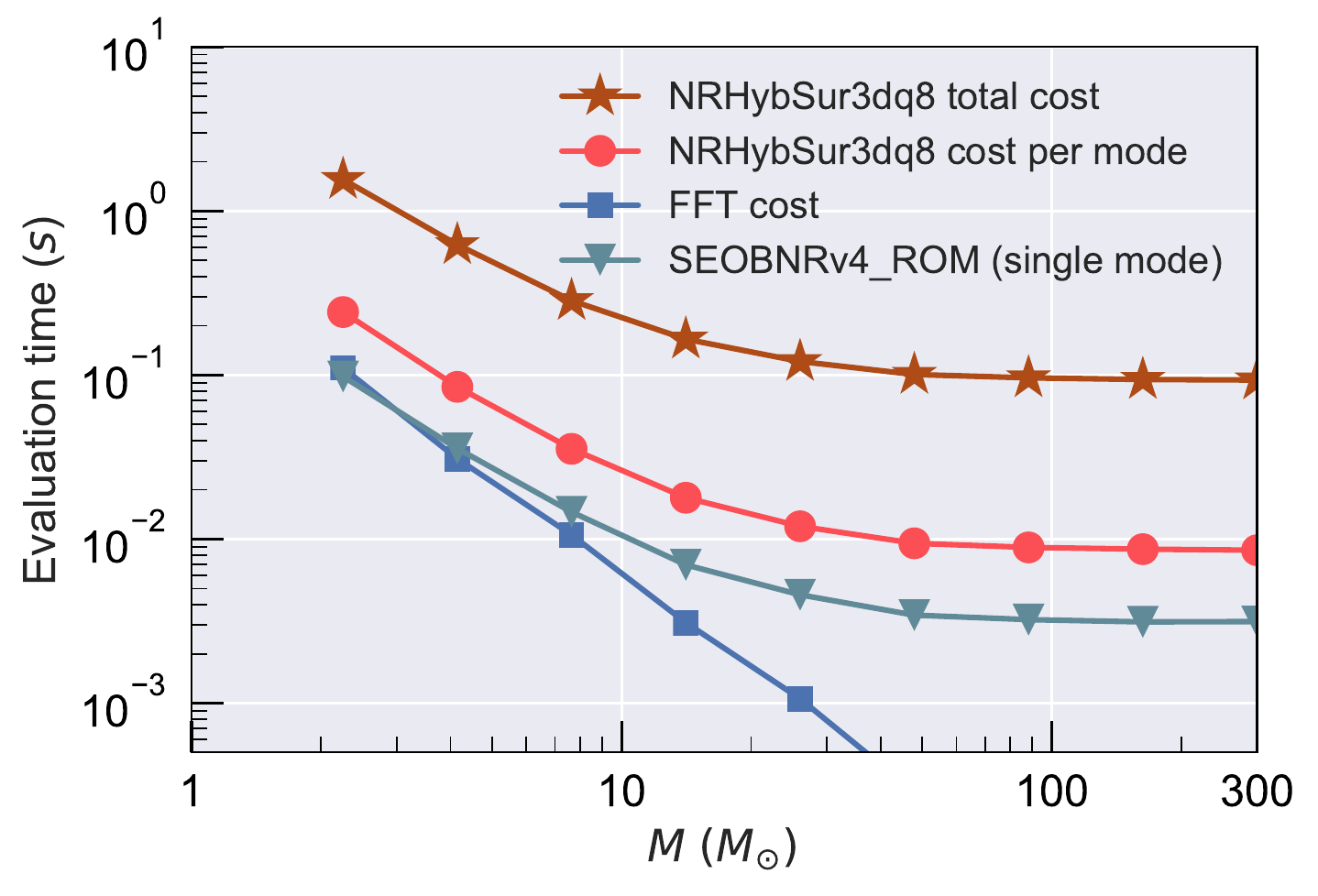}
\caption{Evaluation cost for NRHybSur3dq8 including the cost of an FFT. We show
the cost for evaluating all 11 modes modeled by NRHybSur3dq8, as well as the
cost per mode. The FFT cost is included in both of the above but also
shown separately. We also show the evaluation cost of SEOBNRv4\_ROM which
includes only the $(2,2)$ mode. The evaluation cost is computed by averaging
over 64 points uniformly distributed in the parameter space, $q\leq8$ and
$|\chi_{1z}|,|\chi_{2z}|\leq0.8$.
}
\label{Fig:timing}
\end{figure}

Figure~\ref{Fig:timing} shows the evaluation cost for NRHybSur3dq8, at
different total masses, starting at $20 \text{Hz}$, and using a sampling rate
of $4096 \text{Hz}$. This suggests that NRHybSur3dq8 is fast enough for direct
use in parameter estimation. We also show the evaluation cost per mode.  Note
that the total cost as well the cost per mode in Fig.~\ref{Fig:timing} include
the cost of a Fast Fourier Transform (FFT).  We perform the FFT only once,
after summing over all modes in the time domain. This cost is also shown
separately in Fig.~\ref{Fig:timing}. Finally, we show the evaluation cost of
SEOBNRv4\_ROM~\cite{Bohe:2016gbl}, a Fourier domain Reduced Order Model (ROM)
version of SEOBNRv4.  Note that SEOBNRv4\_ROM models only the $(2,2)$ mode.
Comparing the cost for SEOBNRv4\_ROM to the cost per mode of NRHybSur3dq8
suggests that the evaluation cost of NRHybSur3dq8 can be reduced by a factor of
$\sim2.5$ by building a Fourier domain ROM along the lines of
Ref.~\cite{Purrer:2014fza}.

At low masses, where the waveform is very long, the dominant costs for
NRHybSur3dq8 are due to the temporal interpolation from the sparse domain of
the surrogate to the required time samples, and the FFT. At high masses, where
the waveform is short, the interpolation and FFT are cheap and the dominant
cost for NRHybSur3dq8 is due to the GPR evaluations for the parametric fits.
SEOBNRv4\_ROM instead uses tensor spline interpolation for the parametric
fits~\cite{Bohe:2016gbl}, which accounts for the main difference in the
evaluation cost per mode at high masses.

These tests were performed on a single core on a 3.1\,GHz {Intel Core i5}
processor.  Both NRHybSur3dq8 and SEOBNRv4\_ROM were evaluated using a C
implementation in the LIGO Algorithm Library~\cite{lalsuite}. The Python
implementation of NRHybSur3dq8 in gwsurrogate~\cite{gwsurrogate} is slower than
the C implementation by at most a factor of 2.

\section{Conclusion}
\label{Sec:Conclusion}

We present NRHybSur3dq8, the first NR-based surrogate waveform model that spans
the entire LIGO bandwidth, valid for stellar mass binaries with total masses $M
\geq2.25 M_{\odot}$. This model is trained on 104 NR-PN/EOB hybrid waveforms of
nonprecessing quasicircular BBH systems with mass ratios $q\leq8$, and spin
magnitudes $|\chi_{1z}|,|\chi_{2z}|\leq0.8$.  The parametric fits for this
model are performed using Gaussian process regression. This model includes the
following spin-weighted spherical harmonic modes: $\ell\leq4$ and $(5,5)$, but
not $(4,1)$ or $(4,0)$.  We make our model available publicly through the
easy-to-use Python package \emph{gwsurrogate}~\cite{gwsurrogate}. In addition,
our model is implemented in C with Python wrapping in the LIGO Algorithm
Library~\cite{lalsuite}. We provide an example Python evaluation code at
\cite{SpECSurrogates}.

Through a cross-validation study, we show that the surrogate accurately
reproduces the hybrid waveforms. The mismatch between them is always less than
$\roughly3\times10^{-4}$ for total masses $2.25 M_{\odot}\leq M \leq
300M_{\odot}$. For high masses ($M\gtrsim40M_{\odot}$), where the
merger and ringdown are more prominent, we show roughly a 2 orders of magnitude
improvement over the current state-of-the-art model with nonquadrupole modes,
SEOBNRv4HM~\cite{Cotesta:2018fcv}.

By comparing against several long NR simulations, we show that the errors in
our hybridization procedure are comparable or lower than the resolution error
in current NR simulations. In addition, by comparing against available NR
simulations at higher mass ratios and spins, we show that our model extrapolates
reasonably well outside its training range.  Based on these tests, we are
cautiously optimistic that the surrogate can be used for $q\leq10$ and
$|\chi_{1z}|,|\chi_{2z}|\leq1$, and we leave a more detailed investigation for
future work.

\subsection{Future work}

While our tests of the hybridization procedure are encouraging, long NR
simulations are available only for low mass ratios and low spin magnitudes.
Therefore, we have no means to test hybridization at high mass rations and/or
high spins, where PN is expected to perform poorly.  An improved surrogate
model and refined study of the hybridization errors will require longer
inspiral waveforms with greater coverage of the parameter space.

Another extension of interest is towards larger mass ratios and spin
magnitudes. While the surrogate extrapolates very well when compared to
available simulations at larger mass ratios and spins, no NR simulations are
available with both large mass ratios ($q>8$) and large spins ($\chi>0.8$).
Therefore, our model is untested in that region of
parameter space and it might be
necessary to add training points there. The model could also be extended to
include precession and/or eccentricity, however this is more challenging
because of the enlarged parameter space as well as more complicated
hybridization.

Finally, as mentioned in Sec.~\ref{Subsec:evaluation_cost}, the evaluation time
of NRHybSur3dq8 can likely be reduced by constructing a Fourier domain
ROM~\cite{Purrer:2014fza} of the time-domain model.

We leave these explorations to future work.

\begin{acknowledgments}

We thank Matt Giesler for helping carry out the new SpEC simulations used in
this work. We thank Michael Boyle, Kevin Barkett, Matt Giesler, Yanbei Chen,
and Saul Teukolsky for useful discussions.  We thank Patricia Schmidt for
careful and detailed feedback on an earlier draft of this manuscrtipt. V.V. and
M.S. are supported by the Sherman Fairchild Foundation, and NSF grants
PHY--170212 and PHY--1708213 at Caltech.  L.E.K. acknowledges support from the
Sherman Fairchild Foundation and NSF grant PHY-1606654 at Cornell.  S.E.F is
partially supported by NSF grant PHY-1806665.  This work used the Extreme
Science and Engineering Discovery Environment (XSEDE), which is supported by
National Science Foundation grant number ACI-1548562.  This research is part of
the Blue Waters sustained-petascale computing project, which is supported by
the National Science Foundation (awards OCI-0725070 and ACI-1238993) and the
state of Illinois. Blue Waters is a joint effort of the University of Illinois
at Urbana-Champaign and its National Center for Supercomputing Applications.
Simulations were performed on NSF/NCSA Blue Waters under allocation NSF
PRAC--1713694; on the Wheeler cluster at Caltech, which is supported by the
Sherman Fairchild Foundation and by Caltech; and on XSEDE resources Bridges at
the Pittsburgh Supercomputing Center, Comet at the San Diego Supercomputer
Center, and Stampede and Stampede2 at the Texas Advanced Computing Center,
through allocation TG-PHY990007N.  Computations for building the model were
performed on Wheeler and Stampede2.

\end{acknowledgments}

%
\section*{References}
\bibliography{References/References}

\end{document}